\documentclass{article}
\usepackage[margin=1in]{geometry}
\usepackage[hidelinks]{hyperref}
\usepackage{amsmath}
\usepackage{amssymb}
\usepackage{parskip} 
\usepackage{natbib}
\usepackage{mathrsfs}
\usepackage{pdfpages}
\usepackage{bm}
\usepackage{graphics}
\usepackage{multicol}
\usepackage[bf]{caption}
\usepackage{multirow}
\usepackage{subcaption}
\usepackage{floatrow}
\usepackage{comment}
\usepackage{amsthm}
\usepackage[titletoc]{appendix}
\usepackage{setspace}
\usepackage{mwe}  
\usepackage{color}

\usepackage{authblk}

\usepackage{titlesec}
\usepackage{titlesec}

\begin{document}
\title{\vspace{-1.5cm}Resolving Strain Localization of Brittle and Ductile Deformation in two- and three- dimensions using Graphical Processing Units (GPUs)} %%%%%%%%%%%%

\author[1*,5]{\normalsize Yury Alkhimenkov}
\author[2,5]{\normalsize Lyudmila Khakimova}
\author[3,4,5]{\normalsize Ivan Utkin}
\author[2,5]{\normalsize Yury Podladchikov}

\affil[1]{Department of Civil and Environmental Engineering, Massachusetts Institute of Technology, Cambridge, MA 02139, USA}
\affil[2]{Institute of Earth Sciences, University of Lausanne, Lausanne, Switzerland}
\affil[3]{Laboratory of Hydraulics, Hydrology and Glaciology (VAW), ETH Zurich, Zurich, Switzerland}
\affil[4]{WSL, Birmensdorf, Switzerland}
\affil[5]{Faculty of Mechanics and Mathematics, Lomonosov Moscow State University, Moscow 119991, Russia }
%\\ \linebreak

\date{}
\maketitle

\let\thefootnote\relax
\footnotetext{* yalkhime@mit.edu} %%%%%%%%%%

\begin{abstract}

Shear strain localization refers to the phenomenon of accumulation of material deformation in narrow slip zones. Many materials exhibit strain localization under different spatial and temporal scales, particularly rocks, metals, soils, and concrete. In the Earth's crust, irreversible deformation can occur in brittle as well as in ductile regimes. Modeling of shear zones is essential in the geodynamic framework. Numerical modeling of strain localization remains challenging due to the non-linearity and multi-scale nature of the problem. We develop a numerical approach based on graphical processing units (GPU) to resolve the strain localization in two and three dimensions of a (visco)-hypoelastic-perfectly plastic medium. Our approach allows modeling both the compressible and incompressible visco-elasto-plastic flows. We demonstrate that using sufficiently small strain or strain rate increments, a non-symmetric strain localization pattern is resolved in two- and three-dimensions. We show that elasto-plastic and visco-plastic models yield similar strain localization patterns for material properties relevant to applications in geodynamics. We achieve fast computations using three-dimensional high-resolution models involving more than $500$ million degrees of freedom. We propose a new physics-based approach explaining spontaneous stress drops in a deforming medium. We demonstrate by coupling the geomechanical model with a wave propagation solver that the rapid development of a shear zone in rocks generates seismic signal characteristics for earthquakes.
\end{abstract} %%%%%%%%%

\bigskip

\section{Introduction}

Strain localization refers to the phenomenon of strain accumulation in narrow regions which happens in rocks and in most materials, particularly metals, rocks, soils, and concrete, under different spatial and temporal scales. The physical mechanisms governing strain localization are different in different materials. Usually, localization occurs when the external load reaches a certain threshold. The strength of most geomaterials, particularly rocks, is strongly pressure-dependent, with strength increasing with increasing pressure. In the upper part of the Earth's crust, rocks behave like brittle-elastic material. With increasing depth, growing confining pressure leads to the increase in the strength of the rocks. In the lower parts of the Earth's crust, elevated temperatures activate the ductile regime of deformation. Plastic yielding and strain localization occurs in both brittle and ductile regimes. In this study, we will mostly explore strain localization in rocks, however, the presented methodology is applicable to model the plastic behavior of other materials with similar material properties as well. We here develop efficient numerical algorithms based on High Performance Computing (HPC) and graphical processing units (GPUs) to model strain localization in two- and three- dimensions for applications in geodynamics and earthquake physics.

Numerical modeling is an essential tool to better understand physical processes taking place on Earth \citep{turcotte2002geodynamics}. During the last decades, several three-dimensional geodynamical codes have been developed to model complex physical processes: ASPECT \citep{kronbichler2012high}, Citcom \citep{moresi1996accuracy}, DOUAR \citep{braun2008douar}, FANTOM \citep{thieulot2011fantom}, Fluidity \citep{davies2011fluidity}, I3(E)LVIS \citep{gerya2007robust}, MILAMIN \citep{dabrowski2008milamin}, pTaTin3D \citep{may2015scalable}, PyLith \citep{aagaard2013domain} combined with PETSc \citep{balay2019petsc}, Rhea \citep{burstedde2008scalable}, Slim3D \citep{popov2008slim3d}, TERRA \citep{baumgardner1985three, davies2013hierarchical, wilson2017terra}, STAGYY \citep{tackley1996effects, tackley2008modelling}, Underworld2 \citep{moresi2007computational}. \cite{may2008preconditioned} analyzed several iterative methods for inertialess  flow problems arising in geodynamics. A more complex visco-elastic rheology is needed to adequately model short time-scale \citep{deng1998viscoelastic} and long time-scale processes in geodynamics \citep{farrington2014role, jaquet2016dramatic, olive2016role}. All these codes are mainly implemented using CPUs and are able to model domains including several million elements. However, a new class of numerical algorithms based on GPUs has been recently proposed.

Almost two decades ago, the numerical performance switched from a compute-bound to memory-bound algorithms, this switch being known as ``memory wall''. The most computationally expensive part in modern hardware is reading and writing data from global memory; hundreds of arithmetic operations can be performed per one memory access without limiting performance. Such a revolution accelerated the development of highly parallel devices to overcome limitations related to reading and writing large amounts of data. Graphical processing units or GPUs are the class of highly parallel devices which allow us to execute thousands of programming instructions in a parallel fashion, and memory bandwidth exceeding that of CPUs by orders of magnitude. The algorithms developed for GPUs are very efficient and can tackle three-dimensional problems with a resolution involving trillions of degrees of freedom \citep{alkhimenkov2021resolving, rass2022assessing}. We here list several recently developed GPU-based applications. \cite{omlin2017pore} performed simulations of reactive solitary waves in three-dimensions. \cite{omlin2018simulation, rass2019resolving} used GPUs to resolve viscoelastic deformation coupled to porous fluid flow.  \cite{duretz2019resolving} investigated thermomechanical coupling in two and three dimensions. \cite{rass2020modelling} modeled thermomechanical ice deformation. \cite{alkhimenkov2021resolving, alkhimenkov2021stability} proposed a multi-graphical wave propagation code for poroelastic media resolving over 1.5 billion grid cells in a few seconds. Finally, \cite{rass2022assessing} provided an efficient numerical implementation of pseudo-transient iterative solvers on GPUs using the Julia programming language.

Plasticity has a long history and many scientists contributed into this field. Early studies of strain localization in solids include \citep{hill1998mathematical,hill1958general, kachanov2004fundamentals, rudnicki1975conditions, rice1980note, vardoulakis1978formation, vermeer1984non, sulem1995bifurcation}. The mechanics of faults and earthquakes is given in a monograph by \cite{scholz2019mechanics}. \cite{vermeer1990orientation} analyzed the possible orientations of shear bands in biaxial tests. \cite{forsyth1992finite,buck1993effect} explored the angle of normal faulting. Due to the non-linearity of the problem, numerical modeling of strain localization remains challenging. \cite{cundall1989numerical,cundall1990numerical, poliakov1993initiation, poliakov1994fractal, poliakov1994self} are the early studies addressing strain localization in brittle rocks. The modeling of strain localization in visco-plastic rheology was conducted by \cite{kaus2006initiation}. \cite{kaus2010factors} numerically studied the angle of shear bands considering brittle deformation. \cite{spiegelman2016solvability} explore the algorithms to solve the incompressible equations with visco-plastic rheologies. \cite{glerum2018nonlinear} combined plasticity and nonlinear visco-plastic rheology. Regularization of the strain localizaion thickness was performed by \cite{duretz2019finite} and \cite{de2020viscoplastic}. Geodynamic modeling of frictional plasticity were performed by \cite{duretz2020toward, duretz2021modeling}. \cite{minakov2021elastoplastic} used a two-dimensional elastoplastic model for microearthquake source generation.

The current understanding of the earthquake nucleation is based on studying of the sliding behavior of frictional surfaces. It is believed that the interseismic period is governed by nearly elastic behavior of the crust with periods of anelastic slip \citep{pranger2022rate}. The majority of the modeling studies of earthquake sequences are based on the phenomenological rate- and state-dependent friction law \citep{dieterich1978time, dieterich1979modeling, ruina1983slip}. However, new studies of the earthquake nucleation include additional physical mechanisms, such as plasticity, which might be important for proper modeling of earthquake sequences. The effects of off-fault plasticity in 2-D in-plane dynamic rupture simulations was conducted by \cite{templeton2008off, kaneko2011shallow, gabriel2013source, tong2018simulation, allison2018earthquake}. Off-fault plasticity in three-dimensional dynamic rupture simulations was studied by \cite{wollherr2018off}. \cite{dal2022subduction} presented a 2-D thermomechanical computational framework for simulating earthquake sequences in a non-linear visco-elasto-plastic compressible medium. \cite{uphoff2023discontinuous} used a discontinuous Galerkin method for modeling of earthquake sequences and aseismic slip on multiple faults.

%\hl{IU: add a separate paragraph about earthquake nucleation and mechanisms (Luca Dal Zilio, Casper Pranger). Write about state-of-the art in earthquake modelling and prediction}

%\hl{IU: We take simplest plastic model and through high-resolution and HPC demonstrate the periodic stress drop and characterstic sesimic signal.}

%\cite{scholz2019mechanics}

We here propose a GPU-based numerical implementation of the compressible and incompressible visco-elasto-plastic inertialess equations in two- and three-dimensions. Similarly to \cite{alkhimenkov2021resolving}, the present algorithm relies on the three key ideas: concise numerical implementation, high numerical resolution and high computational efficiency. Concise numerical implementation manifests in short and simple numerical code developed for GPU devices. Spatial and temporal discretization is implemented using a conservative staggered grid numerical scheme \citep{virieux1986p}, equivalent to a variant of a finite-volume method \citep{dormy1995numerical}. The solution of the quasi-static problem is achieved using the matrix free pseudo-transient (relaxation) method \citep{frankel1950convergence, rass2022assessing} which showed its robustness and scalability in large high-performance computing applications. We rely on the perfect plasticity theory and implement a non-associated pressure-dependent Drucker–Prager criterion \citep{drucker1952soil,de2011computational}. The high numerical resolution involving over $130$ million grid cells allows our numerical algorithm to resolve strain localization in two- and three- dimensions. In principle, the resolution is only limited by the available GPU memory and can be up to several billions as it was presented by \cite{alkhimenkov2021resolving, rass2022assessing}. High computational efficiency means that the numerical simulations lasts for several minutes or hours depending on the target numerical resolution. The numerical algorithm is implemented in CUDA C programming language which is suitable for Nvidia GPU devises. The developed algorithm is then applied to model the sequence of stress drops associated with strain localization. Such stress drops can be attributed to spontaneous earthquake nucleation.
%We take simplest plastic model and through high-resolution and HPC demonstrate the periodic stress drop and characterstic sesimic signal.

%\item We demonstrate the similarity in patterns of strain localization between brittle and ductile regimes of deformation;
%\item We propose a new physics-based model explaining spontaneous stress drops in deforming rocks, with potential applications to modelling the nucleation of earthquakes;
%\item We achieve fast numerical simulations in high-resolution model setups involving more than $500$ million degrees of freedom;

The originality of the present study is the following:\\
1. We utilize the simplest pressure-sensitive ideal plasticity model with constant in time and space static friction coefficient.\\
2. We resolve a non-symmetric strain localization pattern in two- and three- dimensions. \\
3. We demonstrate the similarity in patterns of strain localization between elasto-plastic and visco-plastic rheologies for modeling of strain localization. \\
4. We propose a new physics-based approach explaining spontaneous stress drops in deforming rocks, with potential applications to modelling the nucleation of earthquakes. \\
5. We achieve fast computational times using high-resolution models in two- and three-dimensions involving more than $500$ million degrees of freedom.\\

\section{Governing Equations}

We consider a single-phase compressible flow governed by the inertialess equations with visco-elasto-plastic rheology. We use incremental formulation and implement an objective stress rate measure by applying the Jaumann rate. First, we provide the conservation laws and the constitutive relations of the compressible visco-elasto-plastic inertialess equations. A modification leading to the incompressible visco-elasto-plastic inertialess equations is provided in the following section. Then, we explore the Jaumann derivative and non-associated plasticity implemented in the solver. List of principal notation is given in Table \ref{tbl1}.

The model includes equations for conservation of mass, momentum, and closure relations. In this study, we consider only isothermal model setups, thus, we don't include the conservation of energy in the model.

{
%\begin{center}%
\begin{table}[ht] %\centering
\caption{List of Principal Notation}
\centering
\vspace{+1.0 mm}
 \begin{tabular}{| l | l | l |}
  \hline			
  Symbol & Meaning & Unit \\
	\hline
	\hline
   $\sigma_{ij} $   & stress  tensor          & Pa        \\  
   $p$    & pressure             & Pa        \\
   $\tau_{ij} $   & stress deviator           & Pa        \\  
      $\dot{\varepsilon} _{ij}$      & deviatoric strain rate  & 1$/$s        \\
   $v_k $    &  velocity              & m$/$s     \\
   $\rho $    &  density                  & kg$/$m$^3$  \\
   $K$   &  bulk modulus         & Pa        \\%of the grain material
   $G$      &  shear modulus  & Pa        \\ %of the solid grain material
   $\mu_s$      &  shear viscosity  & Pa$\cdot$s        \\ %of the solid grain material
   $t$      & time  & s        \\
  \hline  
 \end{tabular}%\end{center}
 \label{tbl1}
\end{table}
}

\subsection{Conservation laws}

The conservation of mass can be formulated as:

\begin{equation}
    \frac{\partial\rho}{\partial t} + \boldsymbol{\nabla}\cdot(\rho\boldsymbol{v}) = 0 \qquad\textnormal{or}\qquad  \frac{\partial\rho}{\partial t} + \nabla_j(\rho v_j)= 0,
\end{equation}
where $\rho$ is the density, and $\boldsymbol{v}$ is the velocity. The conservation of linear momentum is
\begin{equation}\label{A220}
\begin{aligned}
\pmb{\nabla} \pmb{\cdot} \pmb{\sigma} +  \mathbf{f}= 0 \qquad\textnormal{or}\qquad  \nabla_j \sigma_{ij} + f_i= 0,
\end{aligned}
\end{equation}
where $f_i$ is the body forces. The stress tensor is decomposed into bulk (volumetric) and deviatoric components
\begin{equation}\label{A213}
\begin{aligned}
\pmb{\sigma}  = - p  \mathbf{I}  +\pmb{\tau }   \qquad\textnormal{or}\qquad    \sigma_{ij} = - p \delta_{ij}  +\tau_{ij},
\end{aligned}
\end{equation}
where $\mathbf{I}$ is the second order identity tensor, $\tau_{ij}$ is the stress deviator, $\delta_{ij}$ is the Kronecker delta and $i,j=\overline{1..3}$. Pressure is defined as
\begin{equation}\label{A210}
p= - \frac{1}{3} \mathbf{tr} \pmb{\sigma}  \qquad\textnormal{or}\qquad   p = - \frac{1}{3} \left(\sigma_{xx}+\sigma_{yy}+\sigma_{zz}   \right),
\end{equation}
where $\mathbf{tr}$ is the trace operator. Using equations \eqref{A213}-\eqref{A210}, the conservation of linear momentum can be rewritten as
\begin{equation}\label{A22}
\pmb{\nabla} \pmb{\cdot} ( \pmb{\tau }  - p  \mathbf{I} ) +  \mathbf{f}= 0 \qquad\textnormal{or}\qquad  \nabla_j \left( \tau_{ij} - p \delta_{ij}   \right) + f_i= 0.
\end{equation}
%\subsubsection{Rate formulation}
%
%In the present application, it is convenient to use the rate form of conservation laws. The conservation of linear momentum \eqref{A220} can be expressed in rate form as \citep{rice1980note}
%\begin{equation}\label{A213}
%\nabla_j \left(  \frac{D \sigma_{ij}}{Dt} \right) + f_i= 0,
%\end{equation}
%where ${D} / {D} t$ is the material derivative (explored below).

\subsection{Constitutive equations}

In the compressible case, the compressibility for the density is as follows:
\begin{equation}
    {\rho}\frac{\mathrm{D} p}{\mathrm{D}\rho} = K.
\end{equation}

For the incompressible case, the density is constant:
\begin{equation}
    \rho = \mathrm{const}.
\end{equation}

Combining the continuity equation and the relation between density and pressure yields:
\begin{equation}\label{A21}
\frac{1}{K} \frac{\mathrm{D} p}{\mathrm{D} t}  = -  \nabla_k v_k,
\end{equation}
where $D p / Dt$ is the material derivative (explored below), $\nabla$ is del (or nabla) operator and $k=\overline{1..3}$. 
In the incompressibile case, Eq.~\eqref{A21} reduces to the divergence-free condition:
\begin{equation}\label{A2ssi}
0 = -  \nabla_k v_k.
\end{equation}
%\hl{IU: need separate paragraph about incompressible case}
The strain rate is defined as
\begin{equation}\label{A41}
\dot{\varepsilon} _{ij} = \frac{1}{2} \left( \nabla_i v_j + \nabla_j v_i \right) 
\end{equation}
The rheology is Maxwell visco-elasto-plastic, which is characterized by an additive decomposition of the strain rate into an elastic (volumetric and deviatoric), viscous and plastic components ($\dot{\varepsilon}^{vp} = \dot{\varepsilon}^{vis} + \dot{\varepsilon}^{pl}$)
\begin{equation}\label{A4}
\dot{\varepsilon} _{ij} =  \dot{\varepsilon} _{ij}^{eb} + \dot{\varepsilon} _{ij}^{ed} + \dot{\varepsilon} _{ij}^{vis} + \dot{\varepsilon} _{ij}^{pl},
\end{equation}
where the superscripts $\cdot^{eb}$, $\cdot^{ed}$, $\cdot^{vis}$, $\cdot^{pl}$ denote elastic volumetric (bulk), elastic deviatoric, viscous and plastic parts, respectively. The volumetric (bulk) elastic strain rate is
\begin{equation}\label{A420}
\dot{\varepsilon} _{ij}^{eb} = \frac{1}{3}\nabla_k v_k  \delta_{ij},
\end{equation}
the deviatoric elastic strain rate is
\begin{equation}\label{A42}
\dot{\varepsilon} _{ij}^{ed} = \frac{1}{2 G} \frac{\mathcal{D} \tau_{ij}}{\mathcal{D} t},
\end{equation}
the deviatoric viscous strain rate is
\begin{equation}\label{A43}
\dot{\varepsilon} _{ij}^{vis} = \frac{\tau_{ij}}{2 \mu_s},
\end{equation}
the deviatoric plastic strain rate is
\begin{equation}\label{A44}
\dot{\varepsilon} _{ij}^{pl} = \dot{\lambda} \frac{\partial Q}{\partial \sigma_{ij}},
\end{equation}
where $\dot{\lambda}$ is the plastic multiplier and $Q$ is the plastic flow potential. Combining equations \eqref{A4}-\eqref{A44}, the strain rate can be reformulated as
\begin{equation}\label{A23}
\frac{1}{3}\nabla_k v_k  \delta_{ij} + \frac{1}{2 G} \frac{\mathcal{D} \tau_{ij}}{\mathcal{D} t} + \frac{\tau_{ij}}{2 \mu_s} + \dot{\lambda} \frac{\partial Q}{\partial \sigma_{ij}}= \frac{1}{2} \left( \nabla_i v_j + \nabla_j v_i \right) = \dot{\varepsilon} _{ij}.
\end{equation}

In the limit of infinite viscosity, $\mu_s$, the system of equation becomes the static elasto-plastic model routinely used in solid mechanics \citep{zienkiewicz2005finite}. In the limit of infinite elastic moduli, $K$ and $G$, the system of equation becomes the system of Stokes equations used for creeping low Reynolds number flows in fluid mechanics \citep{zienkiewicz2013finite, happel1983low}.

\subsection{Connection to large strain elasticity theory}

The inelastic response is modeled by hypoelastic constitutive theory. Hypoelasticity corresponds the formulation of the constitutive equations
for stress in terms of objective (frame invariant) stress rates \citep{de2011computational}. The rate evolution law for stress is the following \citep{de2011computational, de2012nonlinear}
\begin{equation}\label{A21q}
\frac{\mathcal{D} \pmb{\sigma}}{\mathcal{D} t} = \mathbf{C}^e : \pmb{\dot{\varepsilon}}^e= \mathbf{C}^e : (\pmb{\dot{\varepsilon}} - \pmb{\dot{\varepsilon}}^{vp}),
\end{equation}
where $\mathcal{D} \tau_{ij} / \mathcal{D} t$ denotes the Jaumann rate of Cauchy stress or simply Jaumann derivative (explored below), $\mathbf{C}^e$ is the tangential elasticity operator, $\pmb{\dot{\varepsilon}}=\pmb{\dot{\varepsilon}}^e+\pmb{\dot{\varepsilon}}^{vp}$ is the strain rate tensor decomposed into elastic $\pmb{\dot{\varepsilon}}^e$ and visco-plastic $\pmb{\dot{\varepsilon}}^{vp}$ parts. For simplicity, we assume that $\mathbf{C}^e$ is the elasticity tensor. We assume that the medium is isotropic and decompose the total stress tensor into the deviatoric component and pressure. The stiffness tensor $C^e_{ijkl}$ can be fully described by the bulk $K$
and shear $G$ moduli:

\begin{equation}\label{eq12}
   C^e_{ijkl} = \left(K - \frac{2}{3} G\right) \delta_{ij} \delta_{kl} + 2 G \left( \frac{1}{2} (\delta_{ik} \delta_{jl} + \delta_{il} \delta_{kj})\right),
\end{equation}
where $ \delta_{ij}$ is the Kronecker delta. The relation \eqref{A21q} reduces to equation \eqref{A23} in the compressible case.

\subsection{Jaumann rate of stress}
The Jaumann rate of Cauchy stress, denoted as $\mathcal{D} \sigma_{ij} / \mathcal{D} t$, is defined as \citep{de2011computational}
\begin{equation}\label{j1}
\frac{\mathcal{D} \sigma_{ij}}{\mathcal{D} t} = \frac{\partial \sigma_{ij}}{\partial t} + v_k \frac{\partial \sigma_{ij}}{\partial x_k} \overbrace{ - \dot{w} _{ik}\sigma_{jk} -  \dot{w} _{jk}\sigma_{ik}}^{-\sigma_{ij}^{\mathcal{R}}},
\end{equation}
where $\dot{w} _{ij}$ is the vorticity tensor defined as
\begin{equation}\label{A5}
\dot{w} _{ij} = \frac{1}{2} \left( \nabla_i v_j - \nabla_j v_i \right).
\end{equation}
and $\sigma_{ij}^{\mathcal{R}}$ is the rotation of the Cauchy stress tensor. The Jaumann derivative consists of stress advection and stress rotation terms. Equation \eqref{j1} can be re-arranged as
\begin{equation}\label{j12}
\frac{\mathcal{D} \sigma_{ij}}{\mathcal{D} t} = \frac{{D} \sigma_{ij}}{{D} t} -{\sigma_{ij}^{\mathcal{R}}},
\end{equation}
where ${D} / \sigma_{ij}{D} t$ is the material derivative:
\begin{equation}\label{j13}
\frac{{D} \sigma_{ij}}{{D} t} = \frac{\partial \sigma_{ij}}{\partial t} + v_k \frac{\partial \sigma_{ij}}{\partial x_k},
\end{equation}
and ${\sigma_{ij}^{\mathcal{R}}}$ is
\begin{equation}\label{j14}
{\sigma_{ij}^{\mathcal{R}}} = \dot{w} _{ik}\sigma_{jk} +  \dot{w} _{jk}\sigma_{ik}
\end{equation}
Expressions for ${\sigma_{ij}^{\mathcal{R}}}$ in two- and three-dimensional configuration are given in \ref{objective_stress}. The Jaumann rate of stress deviator $\tau_{ij}$ can be calculated using the same expressions \eqref{j1}-\eqref{j14} with $\sigma_{ij}$ replaced by $\tau_{ij}$.

% described as  \citep{vermeer1984non,de2011computational,de2012nonlinear, scholz2019mechanics}
\subsection{Plasticity}
%\citep{vermeer1984non, scholz2019mechanics}

Plasticity is implemented using a non-associated pressure-dependent Drucker–Prager criterion \citep{drucker1952soil,de2011computational,de2012nonlinear}. This criterion states that plastic yielding begins
when, the second invariant of the deviatoric stress, $J_2$, and the pressure (mean or hydrostatic
stress), $p$, reach the following condition
\begin{equation}\label{A6}
\sqrt{J_2} -  A p  = B c,
\end{equation}
where $A$ and $B$ are the material parameters determined from experiments, $c$ is the cohesion. In terms of the stress tensor, plastic deformations take place when stresses reach the the yield surface. The yield function $F$ and the plastic potential $Q$ for the Drucker–Prager criterion are defined as 
\begin{equation}\label{A61}
F(\tau,p) = \sqrt{J_2} -  A p  - B c,
\end{equation}
\begin{equation}\label{A6111}
Q(\tau,p) = \sqrt{J_2} - C p.
\end{equation}
In two-dimensional configuration, $A=\sin(\phi)$, $B=\cos(\phi)$ and $C=\sin(\psi)$, where $\phi$ is the angle of internal friction and $\psi \le \phi$ is the dilation angle. Note, that in two-dimensions under plain strain condition, with $\sigma_{zz} = \frac{1}{2} (\sigma_{xx}+\sigma_{yy} )$, Drucker–Prager criterion is eqvivalent to Mohr-Coulumb criterion \citep{templeton2008off}. In three-dimensional configuration, expressions for $A$, $B$ and $C$ are the following
\begin{equation}\label{A611}
A = \frac{6 \sin(\phi)}{ \sqrt{3} (3 - \sin(\phi)},
\end{equation}
\begin{equation}\label{A611}
B = \frac{6 \cos(\phi)}{ \sqrt{3} (3 - \sin(\phi)},
\end{equation}
\begin{equation}\label{A6121}
C = \frac{6 \sin(\psi)}{ \sqrt{3} (3 - \sin(\psi)},
\end{equation}
In two dimensions, the second invariant of the deviatoric stress, $J_2$, is expressed as
\begin{equation}\label{A611}
J_2= \frac{1}{2} \pmb{\tau_{}}: \pmb{\tau_{}} = \frac{1}{2} \tau_{ij} \tau_{ji}  = \frac{1}{2} \left( \tau_{xx}^2+\tau_{yy}^2  \right) + \tau_{xy},
\end{equation}
where the symbol $:$ denotes the double dot product. In three-dimensions, $J_2$ is expressed as
\begin{equation}\label{A611S}
J_2= \frac{1}{2} \pmb{\tau_{}}: \pmb{\tau_{}} = \frac{1}{2} \tau_{ij} \tau_{ji}  = \frac{1}{2} \left( \tau_{xx}^2+\tau_{yy}^2+\tau_{zz}^2  \right) + \tau_{xy}+ \tau_{xz}+ \tau_{yz},
\end{equation}
As long as function $F \leq 0$, the material is in elastic regime. Once $F$ reaches a zero value ($F=0$) plasticity is activated. If the material remains in plastic state ($\partial F / \partial t = 0$) plastic yielding occurs. The present implementation of perfect plasticity requires small temporal increments and is computationally expensive. In our formulation, we assume that the dilation angle $\psi=0$. Thus, the plastic multiplier $\dot{\lambda}$ has a simple form 
\begin{equation}\label{A6121lambda}
\dot{\lambda} = \frac{ A p  + B c}{ \sqrt{J_2} }.
\end{equation}

One way to guaranty spontaneous strain localization is to introduce some form of strain softening that is widely used to ensure formation of the shear bands \citep{lavier1999self, moresi2007incompressible, popov2008slim3d, lemiale2008shear}. However, there are concerns about thermodynamic admissibility of such solutions \citep{duretz2019finite}. Moreover, softening or hardening modulus are small compared to shear modulus and can be neglected as a first-order approximation leading to the ideal plasticity model used in the present study.

%such solutions are not thermodynamically admissible and, therefore, softening is not used in the present study in any form.

%\hl{IU: Need to put the dilation angle to 0 here, otherwise it should pop up in dimensional analysis}

%\hl{IU: Need to add the equation for the plastic multiplier $\lambda$, and discuss the return mapping algorithm.}

%\hl{IU: Body forces not specified}

\subsection{Summary of the governing equations}

Simulations can be conducted by solving the following equations for pressure $p$, stress deviator $\tau_{ij}$ and velocities $v_k$:

\begin{equation}\label{A22S}
\nabla_j \left( \tau_{ij} - p \delta_{ij}   \right) = 0,
\end{equation}

\begin{equation}\label{A21S}
\frac{1}{K} \frac{D p}{D t}  = -  \nabla_k v_k,
\end{equation}

\begin{equation}\label{A23S}
 \frac{1}{2 G} \frac{\mathcal{D} \tau_{ij}}{\mathcal{D} t} + \frac{\tau_{ij}}{2 \mu_s} + \dot{\lambda} \frac{\partial Q}{\partial \sigma_{ij}}= \frac{1}{2} \left( \nabla_i v_j + \nabla_j v_i \right) - \frac{1}{3}\nabla_k v_k  \delta_{ij} .
\end{equation}
In 2D, we have 6 unknowns ($p$, $\tau_{xx}$, $\tau_{yy}$, $\tau_{xy}$, $v_x$, $v_y$) and 6 equations. In 3D, we have 10 unknowns ($p$, $\tau_{xx}$, $\tau_{yy}$, $\tau_{zz}$, $\tau_{xy}$, $\tau_{xz}$, $\tau_{yz}$, $v_x$, $v_y$, $v_z$) and 10 equations, respectively.

\subsubsection{Nondimensionalization}

We choose the following dimensionally independent scales: length $l^* = L_x$, time $t^* = 1/a$ and pressure $p^* = G_0$. $L_x$ is the size of the computational domain in $x$-dimension, $a$ is the background strain rate. After rescaling, the model is defined by 4 non-dimensional numbers summarized in the Table~\ref{tbl2}.

The deformation occurs at times inversely proportional to the background strain rate $a_0$ at time $t=0$. Deborah number $De$ characterises the ratio between and the Maxwell relaxation time $\eta/G_0$ and the characteristic time of deformation $t^*$. Thus, if $De >> 1$, the deformation occurs at much shorter time scales than stress relaxation, corresponding to brittle-like behavior (elastic domain). Conversely, if $De << 1$, the stress relaxation is much faster then the deformation, corresponding to the fluid-like, or ductile, behavior (viscous domain). In the ductile regime, high strain rates are necessary to build up stresses sufficient for plastic deformation. $\nu$ is the Poisson's ratio characterising relative effect of compressibility. $r$ characterises the ratio between cohesion $c$ and the pressure scale $p^*$.

%The nondimensionalization is performed by using a new methodology. We divide all our variables into three categories: dimensionally independent, non-dimensional and dimensionally dependent. In our modeling study, dimensionally independent parameters are: the size of the model in $x$-dimension $L_x$, the shear modulus $G$, background velocity increment $a$. Non-dimensional parameters are given in Table \ref{tbl1}. All other parameters are dimensionally dependent, for example, the $y$-dimension of the model is $L_y=L_y\_L_x \cdot L_x$,the bulk modulus is $K= K\_G \cdot G$, etc. We kept the dimensional equations in the code but with properly designed input parameters divided into these three categories. This allows us to set all dimensionally independent parameters equal to 1 ($L_x=1$, $G=1$, $a=1$) and re-scale the input parameters via non-dimensional parameters.

{
%\begin{center}%
\begin{table}[ht] %\centering
\caption{Independent scales}
\centering
\vspace{+1.0 mm}
 \begin{tabular}{| l | l | }
  \hline			
  Scale & Meaning  \\
	\hline
	\hline
   $l^*=L_x$   & Domain size in $x$-direction  \\  
   $t^* = \cfrac{1}{a}$    & time    \\
   $p^* = G_0$    & pressure           \\  
  \hline  
 \end{tabular}%\end{center}
 \label{tbl1}
\end{table}
}

{
%\begin{center}%
\begin{table}[ht] %\centering
\caption{Dimensionless parameters}
\centering
\vspace{+1.0 mm}
 \begin{tabular}{| l | l | }
  \hline			
  Dimensionless number & Meaning  \\
	\hline
	\hline
    $De     = a_0\cfrac{\eta}{G_0}$ & Deborah \\
    $\nu    = \cfrac{3K-2G_0}{2(3K-G_0)}$ & Poisson's ratio \\
    $r      = \cfrac{c_0}{G_0}$ & cohesion to pressure scale \\
    $\gamma = \cfrac{\mathrm{d}a}{\mathrm{d}t} {t^*}^2$ & increment of strain rate \\
  \hline  
 \end{tabular}%\end{center}
 \label{tbl2}
\end{table}
}

\clearpage

%First, we consider dimensionaly independent parameters.

%Dimensionally dependent parameters are the Deborah number $De$, friction angle $\phi$, $\sigma_0$.

%As a result, we solve the same equations as we would in dimensional case but with input parameters categorized in three categories.

%which is equivalent to a finite volume approach \citep{dormy1995numerical}.  We provide the Matlab, symbolic Maple and GPU CUDA C routines to reproduce the main presented results. These routines are available for download from Bitbucket at \url{https://bitbucket.org/yalkhimenkov/fastbiot_gpu3d_v1.0} (last access: 8 February 2021). The routines archive (v1.0) \citep{alkhimenkov_yury_2021_4519367} is available from a permanent DOI repository (Zenodo) at \url{http://doi.org/10.5281/zenodo.4519367} (last access: 8 February 2021).\\

%The novelties of the present article are summarized as following:%\\

%We develop a tool for practical applications delivering results on the order of seconds or minutes of computational time. 
%1. We present 1.\\

%2. We perform 2. \\
%dispersion analysis to understand the behavior of the dimensionless elastodynamic Biot's equations

\section{Numerical solution strategy}

\subsection{Discretization}

 The model domain is discretized using a regular time-space grid. A solution of the system \eqref{A22S}-\eqref{A23S} is performed using a conservative staggered space-time grid discretization \citep{virieux1982dynamic, virieux1986p}. In this approach, fluxes are calculated at the cell boundaries whereas field variables are located at either at cell centers or corners. As a result, a conservative scheme space-time numerical scheme is achieved, which is equivalent to a finite volume method \citep{dormy1995numerical, leveque1992numerical}. The current configuration of the staggered space-time grid discretization is provided in \cite{alkhimenkov2021resolving}; a review of the explicit staggered grid approach is given by \cite{moczo2007finite}.

\subsection{Accelerated pseudo-transient method}

An iterative matrix-free pseudo-transient method \citep{frankel1950convergence, rass2022assessing} is used to obtain the solution of the quasi-static problem. The pseudo-transient method is a physics-inspired iterative method that solves the quasi-static equations by adding a pseudo-time derivative, through which the steady-state solution is progressively achieved via a pseudo-time stepping. In other words, the quasi-static solution is an attractor of the dynamic equation (with inertia) with a pseudo-time derivative. This method is also known as the relaxation method \citep{frankel1950convergence}.

A steady-state solution of the given system of equations \eqref{A22S}-\eqref{A23S} can be performed by converting the equations into a pseudo-transient formulation in two steps: (i) adding inertia term with a "pseudo" time step $\widetilde{t}$ into the momentum equation and (ii) adding Maxwell type viscous rheology into the constitutive equation. The pseudo-transient version of the momentum equation is
\begin{equation}\label{A22P}
\widetilde{\rho} \frac{\partial v_i}{\partial \widetilde{t}} = \nabla_j \left( \tau_{ij} - p \delta_{ij}   \right) + f_i
\end{equation}
In compressible case, the pseudo-transient version of the equation for pressure \eqref{A21} becomes:
\begin{equation}\label{A2ssP}
\frac{1}{\widetilde{K}} \frac{\partial p}{\partial \widetilde{t}} +  \frac{1}{K} \frac{ p - \hat{p}}{\Delta t}  = -  \nabla_k v_k
\end{equation}
In incompressible case, the pseudo-transient version of equation \eqref{A2ssi} becomes:
\begin{equation}\label{A2ssP2}
\frac{1}{\widetilde{K}} \frac{\partial p}{\partial \widetilde{t}} = -  \nabla_k v_k
\end{equation}

For the stress deviator the corresponding equation is
\begin{equation}\label{A23ssP}
\frac{1}{2 \widetilde{G}} \frac{\partial{} \tau_{ij}}{\partial{} \widetilde{t}}   +    \frac{1}{2 G} \frac{ \tau_{ij} - \hat{\tau}_{ij}}{\Delta t} + \frac{\tau_{ij}}{2 \mu_s} = - \dot{\lambda} \frac{\partial Q}{\partial \sigma_{ij}} + \frac{1}{2} \left( \nabla_i v_j + \nabla_j v_i - \frac{2}{3}\nabla_k v_k \right).
\end{equation}
The only difference between the pseudo-transient versions of compressible and incompressible equations is the equation for pressure (equation \eqref{A2ssP} or \eqref{A2ssP2}), all other equations remain the same.

In \eqref{A22P}, \eqref{A2ssP}, \eqref{A23ssP}, the quantities $\widetilde{K}$, $\widetilde{G}$ and $\widetilde{\rho}$ are to be determined numerical parameters. \cite{rass2022assessing} showed that the optimal values of the numerical parameters for the incompressible visco-elastic inertialess equation are
\begin{equation}\label{PT1}
\widetilde{\rho} = \widetilde{Re} \frac{\mu_s}{\widetilde{V} L},
\end{equation}
\begin{equation}\label{PT2}
\widetilde{G} = \frac{\widetilde{\rho}\widetilde{V}^2 }{r + \frac{4}{3}},
\end{equation}
\begin{equation}\label{PT23}
\widetilde{K} = r \widetilde{G},
\end{equation}
where $r = \widetilde{K} / \widetilde{G}$, $\widetilde{Re}=\widetilde{\rho} V_p L/ \mu_s$ is the numerical Reynolds number and the primary, or P-wave, velocity is
\begin{equation}\label{PT24}
V_p = \sqrt{\frac{\widetilde{K} + \frac{4}{3} \widetilde{G}}{ \widetilde{\rho} }}.
\end{equation}
 
Only two numerical parameters, $\widetilde{Re}$ and $r$, control the convergence of the pseud-transient method. A one-dimensional dispersion analysis for the incompressible visco-elastic equation provide us with the following optimal values \citep{rass2022assessing}
\begin{equation}\label{PT24q}
\widetilde{Re}_{opt} = \frac{3 \sqrt{10}}{2} \pi,
\end{equation}
\begin{equation}\label{PT24qq}
r_{opt} = \frac{1}{2}  .
\end{equation}
Numerical tests in two- and three- dimensions show that the derived values for $Re_{opt}$ and $r_{opt}$ remain valid for the compressible visco-elastic equation as well.

\subsection{Return mapping algorithm for plasticity}

In the return mapping algorithm, once the trial deviatoric stress reaches the value of $\overline{\overline{c}} = A p  + B c$, the deviatoric updated stress is obtained by simply scaling down the trial deviatoric stress by a factor that depends on $\Delta{\lambda}$ and is defined by equation \eqref{A6121lambda}. In two- and three- dimensional large scale numerical simulations involving more than 1 million grid cells, we used the concept of the "frozen" plasticity. If after $16 \times n_x$ iterations ($n_x$ - number of cells in $x$-dimension) for a given time step (increment), the relative error $\epsilon_\mathrm{rel}$ (eq. \eqref{Convergence2}) is larger than $10^{-4}$, we stop updating $\dot{\lambda}$ (i.e., we "freeze" plastic multiplier) and iterate until the solution is converged to the desired tolerance $\epsilon_\mathrm{rel}$.

%\subsection{Accelerated pseudo-transient method}

\subsection{The CUDA C implementation using graphical processing units (GPUs)}

\begin{figure}
\centering
\includegraphics[width=0.99\textwidth]{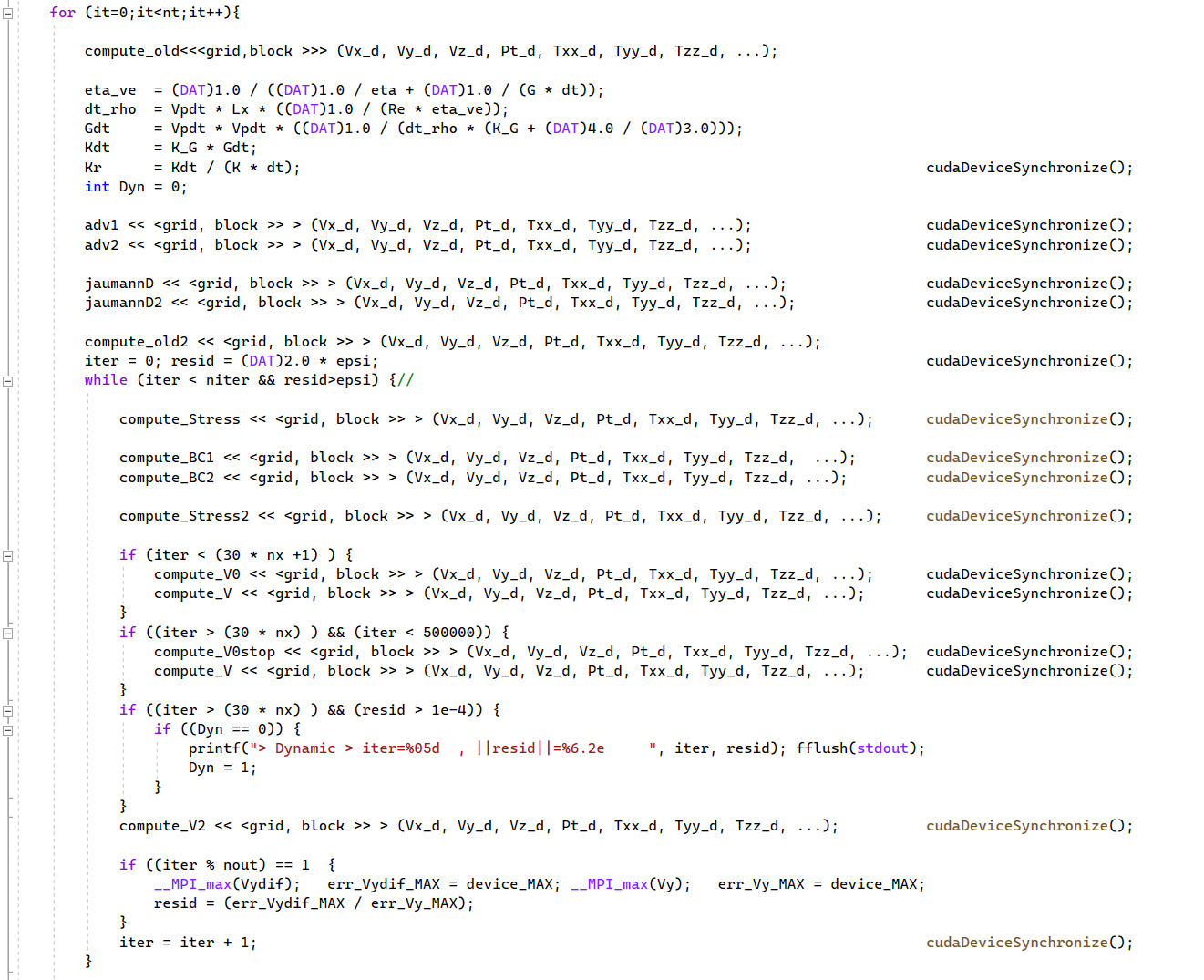}
\caption{Time loop computations of compressible visco-elasto-plastic solver for a single GPU CUDA C code implementation. \texttt{DAT} is a shortcut for the double precision. Kernels \texttt{adv1} and \texttt{adv1} denote the advection. Kernels \texttt{jaumannD} and \texttt{jaumannD2} correspond to the stress rotation. Kernels \texttt{compute\_old} and \texttt{compute\_old2} stand for saving and exchanging values between current and previous time step variables. Kernels \texttt{compute\_Stress} and \texttt{compute\_Stress2} correspond to the update of all stresses. Kernels \texttt{compute\_BC1} and \texttt{compute\_BC2} stand for boundary conditions. Kernels \texttt{compute\_V0} and \texttt{compute\_V} stand for plasticity update (kernel \texttt{compute\_V0stop} corresponds to the "frozen" plasticity). Kernel \texttt{compute\_V2} corresponds to the update of velocities. } 
\label{Code}%
\end{figure}

We use the CUDA C language to implement the numerical solver. The solver consists of two time loops: the first time loop belongs to physical time $t$ whereas the second time loop belongs to the pseudo-time $\widetilde{t}$ integration to achieve a steady-state solution at each discrete physical time step. Figure \ref{Code} shows the time loop computations of compressible visco-elasto-plastic solver for a single GPU CUDA C code implementation. The time loops call several kernels (GPU functions) to sequentially solve the system of equations \eqref{A22S}-\eqref{A23S}. 

We calculated our results using several computing systems. The code prototyping was developed on a laptop hosting $13^{th}$ Gen Intel Core i9-13900HX CPU (64GB RAM) and NVIDIA GeForce RTX 4090 (16GB) laptop GPU. Large scale two- and three- dimensional simulations were conducted on an NVIDIA DGX-1 - like node hosting 4 NVIDIA Ampere A100 (80GB) GPUs and AMD EPYC 7742 (512GB RAM) Server Processor.
%\citep{simo2006computational} computational inelasticity

\section{Modeling Results}

\subsection{Model Configuration and Boundary conditions}

The visco-elasto-plastic solver is represented by the system of equations \eqref{A22S}-\eqref{A23S}. The computational domain is a cube (or square in 2D) with dimensions $x,y,z \in [0,L_x]\times[0,L_y]\times[0,L_z]$ (in 2D $x,y \in [0,L_x]\times[0,L_y]$). All simulations presented in this study have been performed using a simple initial model configuration. The pure shear boundary conditions are applied by prescribing velocities at all boundaries. In two-dimensions, we set
\begin{equation}\label{A7}
v_x = a x 
\end{equation}
and
\begin{equation}\label{A8}
v_y = - a y, 
\end{equation}
In three-dimensions, we prescribe
\begin{equation}\label{A7M}
v_x = a x ,
\end{equation}
\begin{equation}\label{A8M} % \frac{a}{2}
v_y = - a y, 
\end{equation}
and
\begin{equation}\label{A9M}
v_z = 0, 
\end{equation}
which corresponds to the extension in x-dimension and compression in y- dimension. In the brittle regime (elastic domain), we impose loading increments applied to the strain components. In the ductile regime (viscous domain), we impose velocity increments. At all boundaries, free-slip boundary conditions are implemented.

We introduce the non-dimensional shear modulus and non-dimensional cohesion:

\begin{align}
    \bar{G} &= G/G_0, \\
    \bar{c} &= c/c_0.
\end{align}

In all simulations, except for the first series of computations (section "Symmetry versus non-symmetry"), the following initial conditions and heterogeneities are employed. All initial conditions (pressure, deviatoric stress and strain) are set equal to zero. We set anomalies to the non-dimensional shear modulus $\bar{G}$ and cohesion $\bar{c}$. The shear modulus $\bar{G}$ is represented by a Gaussian distribution with the lowest value ($\bar{G}=1$) attributed to the center of the model, $\bar{G}$ gradually increases towards the walls of the model reaching the highest value of $\bar{G}=1.2$. The non-dimensional cohesion $\bar{c}$ has a similar Gaussian distribution with the lowest value ($\bar{c}=0.2\cdot 10^{-2}$) in the center of the model; in addition, a sharp jump to the higher value ($\bar{c}=0.4\cdot 10^{-2}$) is introduced near the walls of the model to eliminate possible boundary effects. The angle of internal friction is $\phi=30^\circ$ in all computations. %In all our simulations, we set $\psi=0$ for simplicity. 

\subsection{Convergence study}

We use $L^\infty$ metric space for measuring the error:
\begin{equation}\label{Convergence1}
L^\infty(\boldsymbol{V}) = \max_i{|V_i|}
\end{equation}
The relative error is calculated as
\begin{equation}\label{Convergence2}
\epsilon_\mathrm{rel} = \frac{\max_i\{|V_i - V_i^{it}|\}}{\max_i\{|V_i|\}}
\end{equation}
and the residual equation error is
\begin{equation}\label{Convergence3}
\epsilon_\mathrm{abs} = \max_i\{|-\nabla p_i + \nabla\cdot \boldsymbol{\tau}_i|\}.
\end{equation}

\begin{figure}
\centering
\includegraphics[width=0.99\textwidth]{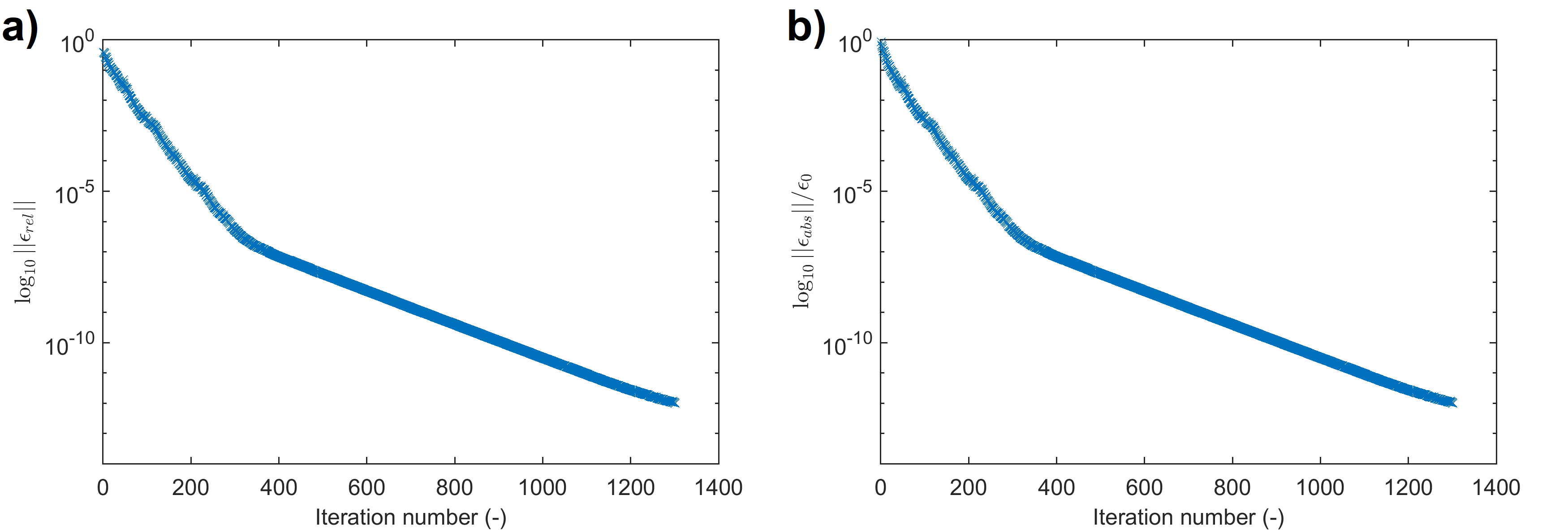}
\caption{Evolution of the relative $L^\infty$ norm of residuals throughout the localization phase. The model configuration was computed using $1535 \times 767$ cells. Panel (a) shows the relative residual error $\epsilon_\mathrm{rel}$ for a (typical) $10^{th}$ strain increment. Panel (b) shows the absolute residual error $\epsilon_\mathrm{abs}$ for a (typical) $10^{th}$ strain increment.} 
\label{Iteration1}%
\end{figure}

Figure \ref{Iteration1} shows a typical evolution of the relative $L^\infty$ norm of residuals throughout the localization phase. The relative residual error $\epsilon_\mathrm{rel}$ for a (typical) $10^{th}$ strain increment of a simulation involving $1535 \times 767$ cells is shown in Figure \ref{Iteration1}a. The absolute residual error $\epsilon_\mathrm{abs}$ for the same simulation is shown in Figure \ref{Iteration1}b.

\subsection{Symmetry versus non-symmetry}

\begin{figure}
\centering
\includegraphics[width=0.99\textwidth]{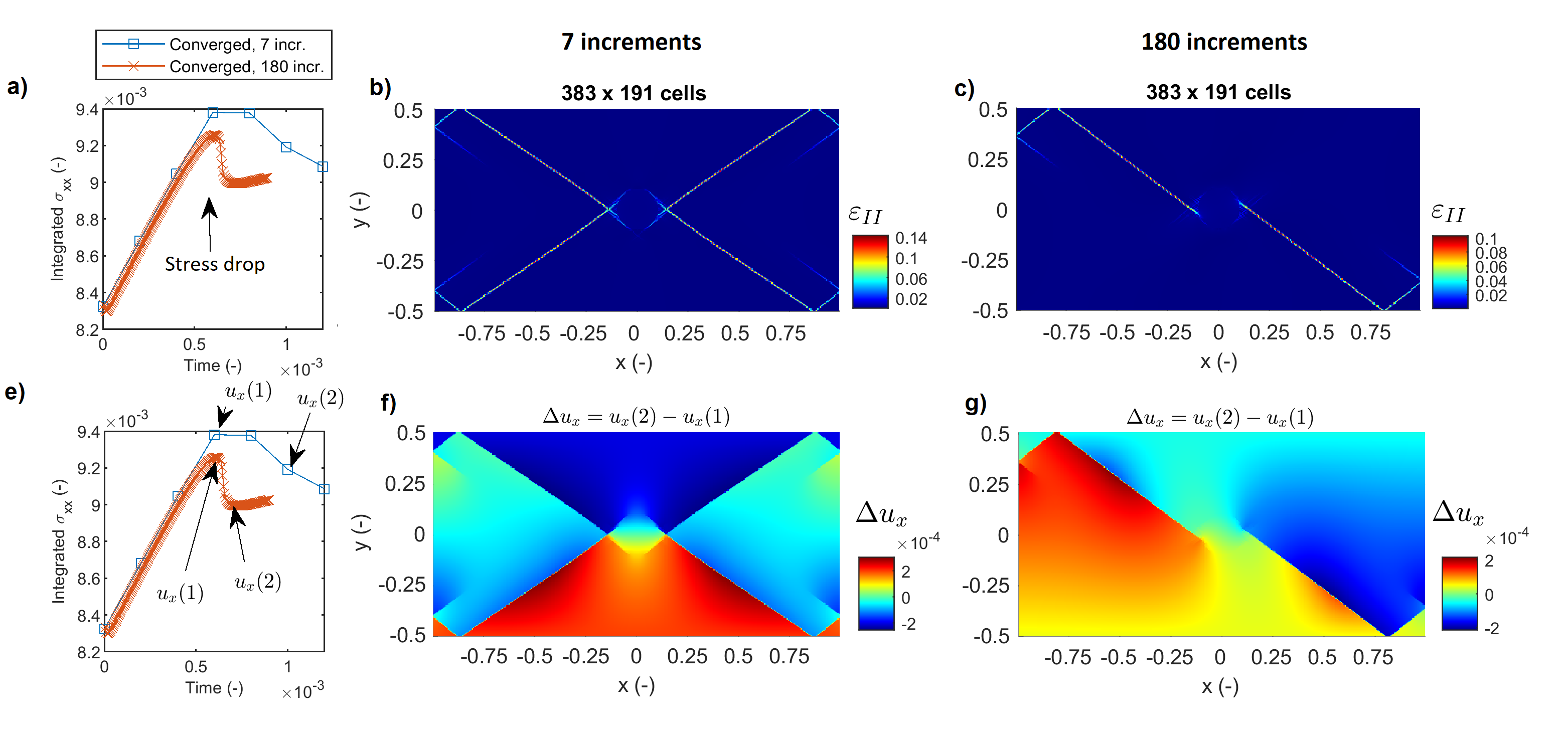}
\caption{Fully converged solutions without Jaumann derivative. Integrated stress over one vertical line segment versus the number of increments (panels ae), the second invariant of the accumulated strain $\varepsilon_{II}$ for different temporal and spatial resolutions (panels bc) and the incremental displacement $\Delta u$ corresponding to the stress drop (panels fg). Note the non-symmetric pattern of $\varepsilon_{II}$ and $\Delta u$ corresponding the solution with small increments.} 
\label{New_nSS}%
\end{figure}

The first series of computations are carried using compressible visco-elasto-plastic equations. The Deborah number is set to $De=10^6$ which corresponds to the brittle domain (elastic limit). We set a circular pressure anomaly to the center of the model ($p=\bar{c}/2$ inside the perturbation, $p=0$ outside the perturbation). We set the shear modulus $\bar{G}=1$ and the cohesion $\bar{c}=0.2\cdot 10^{-2}$ in the entire model domain. Figure \ref{New_nSS}ae shows the integrated stress over one vertical line segment versus time. Figure \ref{New_nSS}bc shows the second invariant of the accumulated strain $\varepsilon_{II}$ for two different temporal resolutions. 

\begin{figure}
\centering
\includegraphics[width=0.55\textwidth]{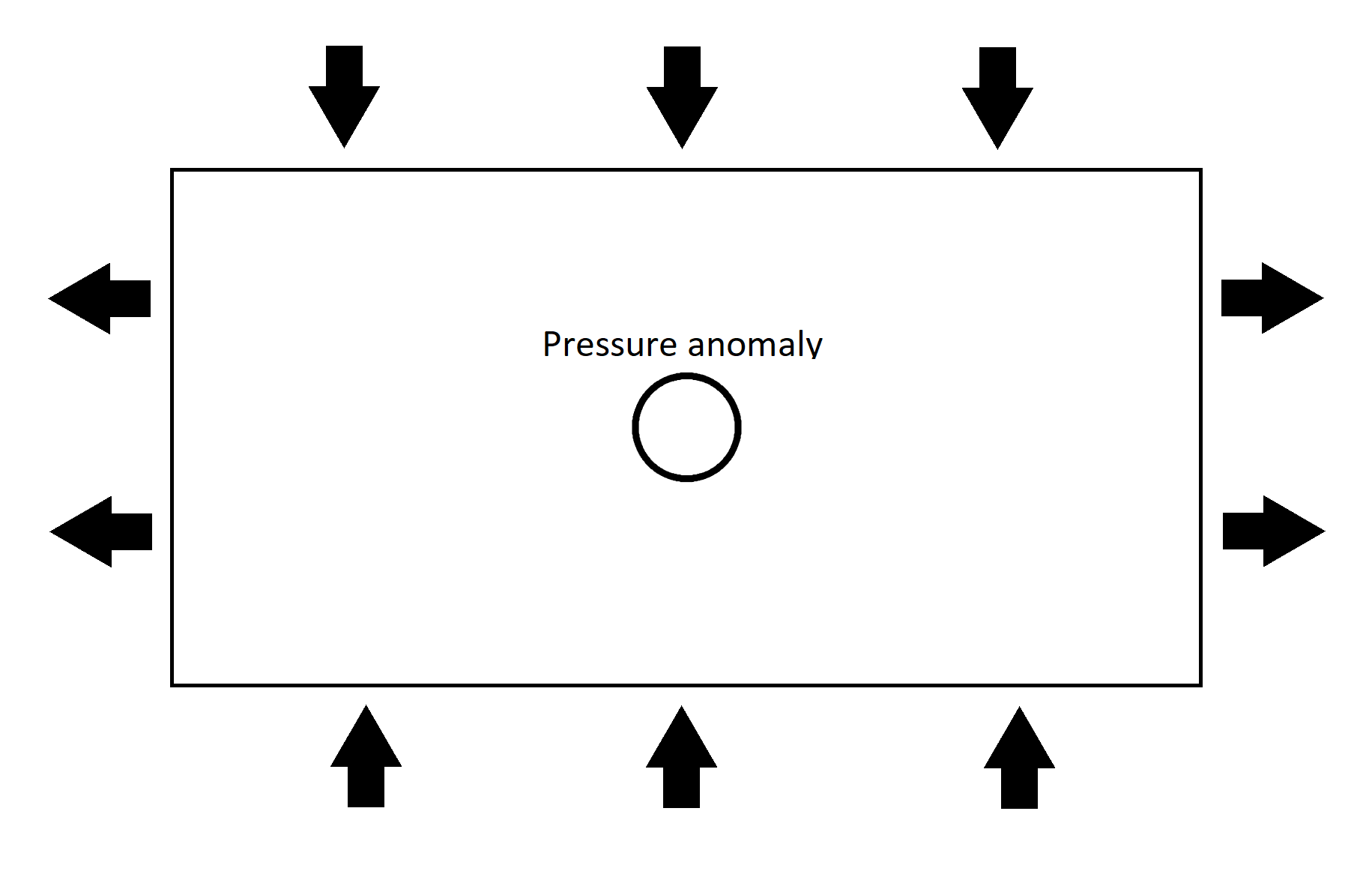}
\caption{Model configuration. The arrows indicate the pure shear boundary condition which is applied at the model boundaries.} 
\label{MODEL12}%
\end{figure}

First, we run two low-resolution simulations of $383\times191$ grid cells. Note that in this example, for simplicity, we exclude the Jaumann derivative. In the first run, only 7 increments in time were performed (with a large time step) (Figure \ref{New_nSS}b). At each time increment we fully converged to the desired relative error $\epsilon_\mathrm{rel}$ of $10^{-12}$ (equation \eqref{Convergence2}). Four shear bands evolve starting from the pressure anomaly and grow towards the walls of the model. An incorrect symmetric strain localization pattern is clearly visible. Contrary, Figure \ref{New_nSS}c corresponds to the calculation of 180 increments in time (with a small time step), at each increment the iterations converged to the desired relative error $\epsilon_\mathrm{rel}$ of $10^{-12}$. Two shear bands develop starting from the pressure anomaly and grow towards the walls of the model, than reflect. The resulting solution exhibits a correct non-symmetric strain localization pattern. Note, that the non-symmetric solution corresponds to the sharper stress drop (and lower stress) than the symmetric solution (Figure \ref{New_nSS}a). To further investigate the difference, we plot the displacement increments corresponding to the stress drop $\Delta u_x = u_x(2) - u_x(1)$ for both, symmetric and non-symmetric solutions (Figure \ref{New_nSS}fg). Sharp discontinuances observed in both solutions. Note that we observe stress drops in deforming medium even without material strain softening or dynamic reduction of the static friction coefficient.

\begin{figure}
\centering
\includegraphics[width=0.7\textwidth]{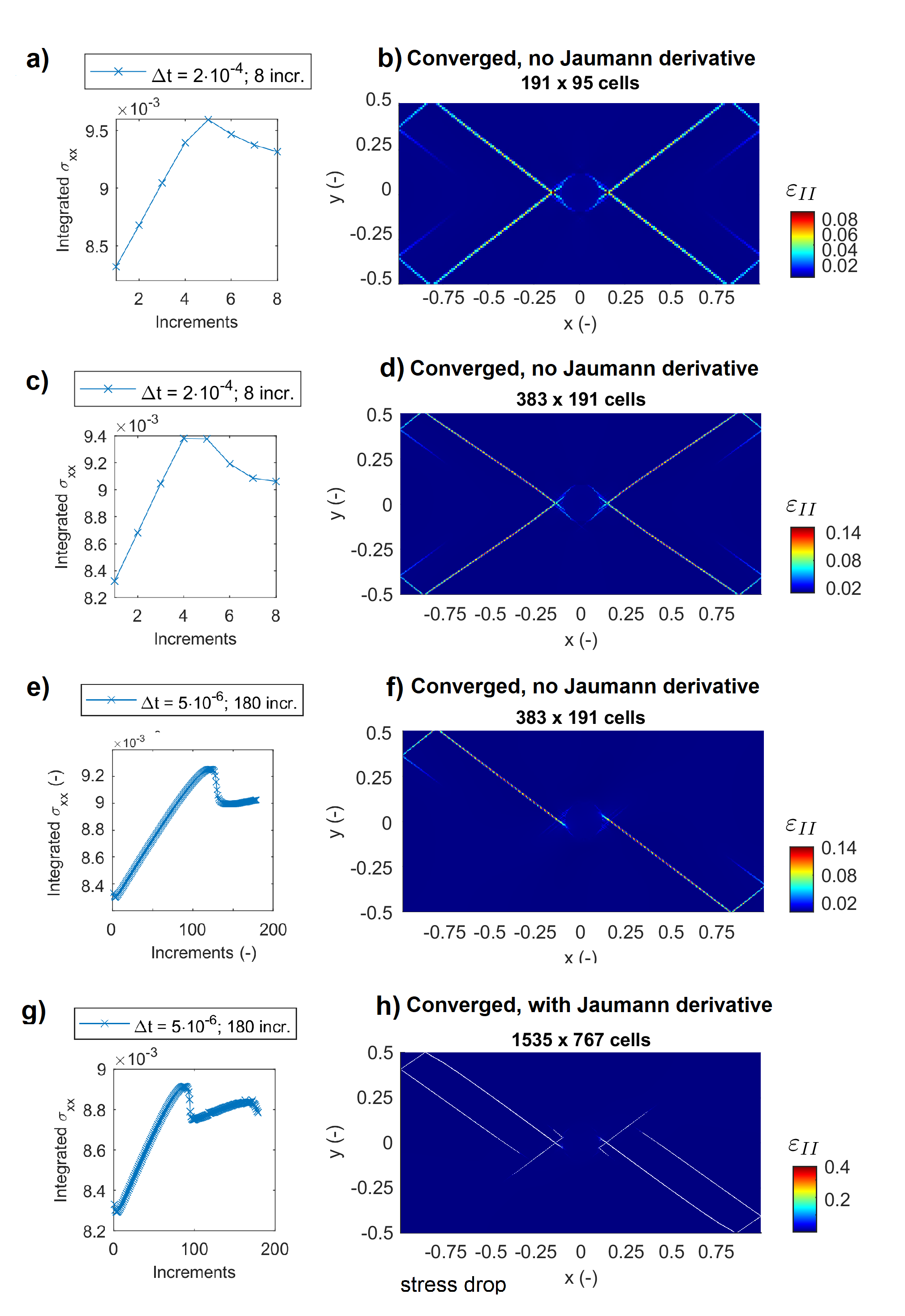}
\caption{Integrated stress over one vertical line segment versus the number of increments (panels acegi) and the second invariant of the accumulated strain $\varepsilon_{II}$ (panels bdfhj) for different temporal and spatial resolutions.} 
\label{Sym_191v3S}%
\end{figure}

In this set of experiments, we investigate both the temporal and spatial resolutions, as well as, the Jaumann derivative. Figure \ref{Sym_191v3S} shows the integrated stress over one vertical line segment versus the number of increments and the second invariant of the accumulated strain $\varepsilon_{II}$ for different temporal and spatial resolutions. Two to four shear bands develop starting from the pressure anomaly (Figure \ref{Sym_191v3S}bdfh). First, we run a low-resolution simulation of $191\times95$ grid cells, 8 increments in total (Figure \ref{Sym_191v3S}b). At each time increment we fully converged to the desired relative error $\epsilon_\mathrm{rel}$ of $10^{-12}$. A symmetric strain localization pattern is clearly visible. Then, we run a simulation for the resolution of $383\times191$ grid cells, 8 increments in total (Figure \ref{Sym_191v3S}d). At each time increment we fully converged to the desired relative error $\epsilon_\mathrm{rel}$ of $10^{-12}$. Again, a symmetric strain localization pattern can be observed. For the next simulation, we increased the temporal resolution by increasing the number of loading steps (i.e., by decreasing the time step) and performed 180 increments. At each increment the simulation converged to the desired relative error $\epsilon_\mathrm{rel}$ of $10^{-12}$. As a result, a correct non-symmetric strain localization pattern is visible (Figure \ref{Sym_191v3S}f). Note, that the stress drop is also visible in Figure \ref{Sym_191v3S}e compared to previous simulations in Figures \ref{Sym_191v3S}ac. In a similar way, we performed a simulation of $1535\times767$ grid cells and kept the same temporal resolution. Again, a correct non-symmetric strain localization pattern (Figure \ref{Sym_191v3S}h) as well as the stress drop (Figure \ref{Sym_191v3S}g) are visible. 

%In the last simulation in this set of tests, we increased the temporal resolution and performed 180 increments in time (Figure \ref{Sym_191v3S}j). A sharp stress drop can be observed (Figure \ref{Sym_191v3S}i).

\begin{figure}
\centering
\includegraphics[width=0.99\textwidth]{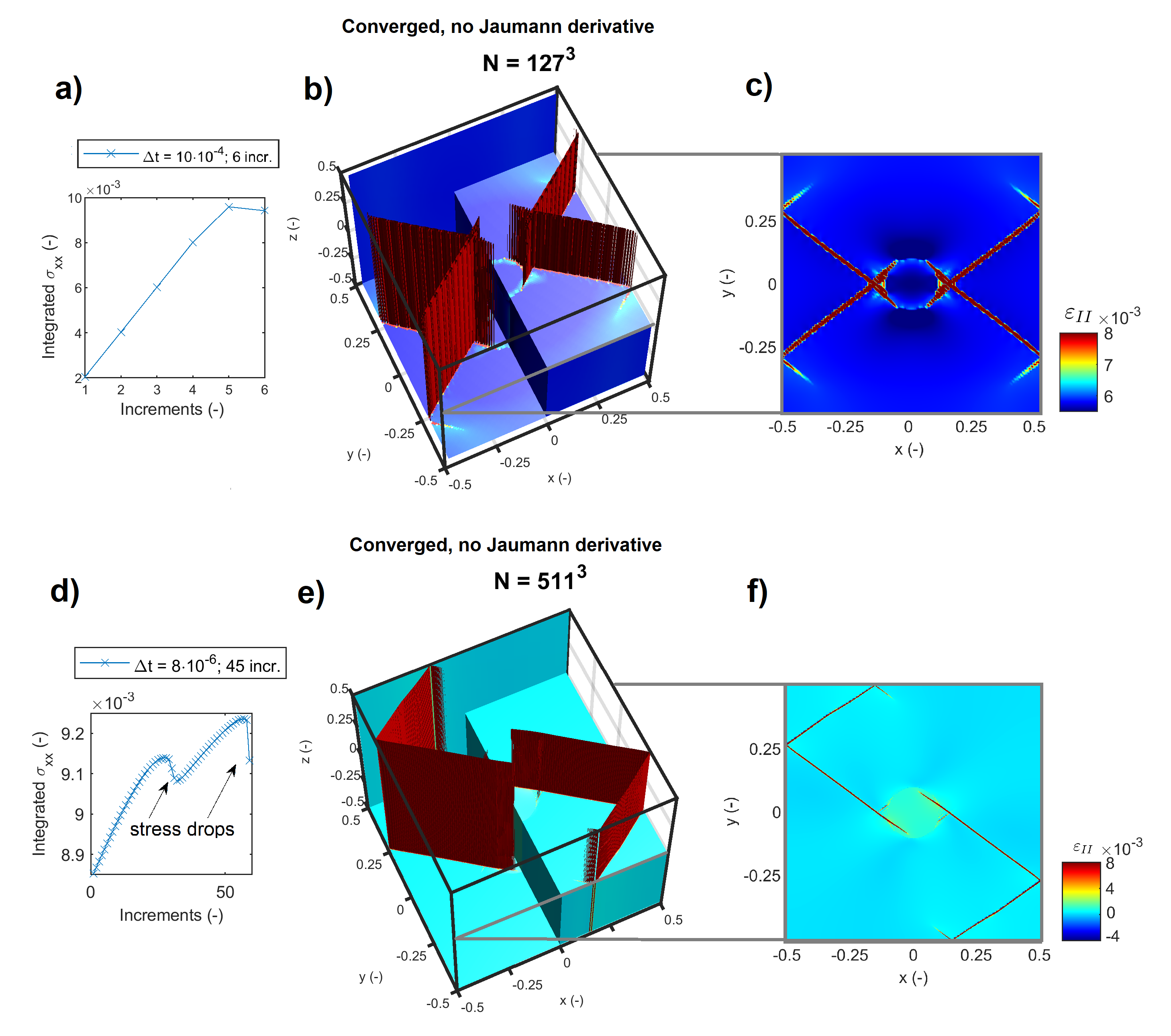}
\caption{Three-dimensional simulation: integrated stress over one vertical line segment versus the number of increments (panels ad) and the second invariant of the accumulated strain $\varepsilon_{II}$ (panels bcef).} 
\label{3D}%
\end{figure}

\clearpage

\subsection{Incompressible elastic and viscous limits}

\begin{figure}
\centering
\includegraphics[width=0.7\textwidth]{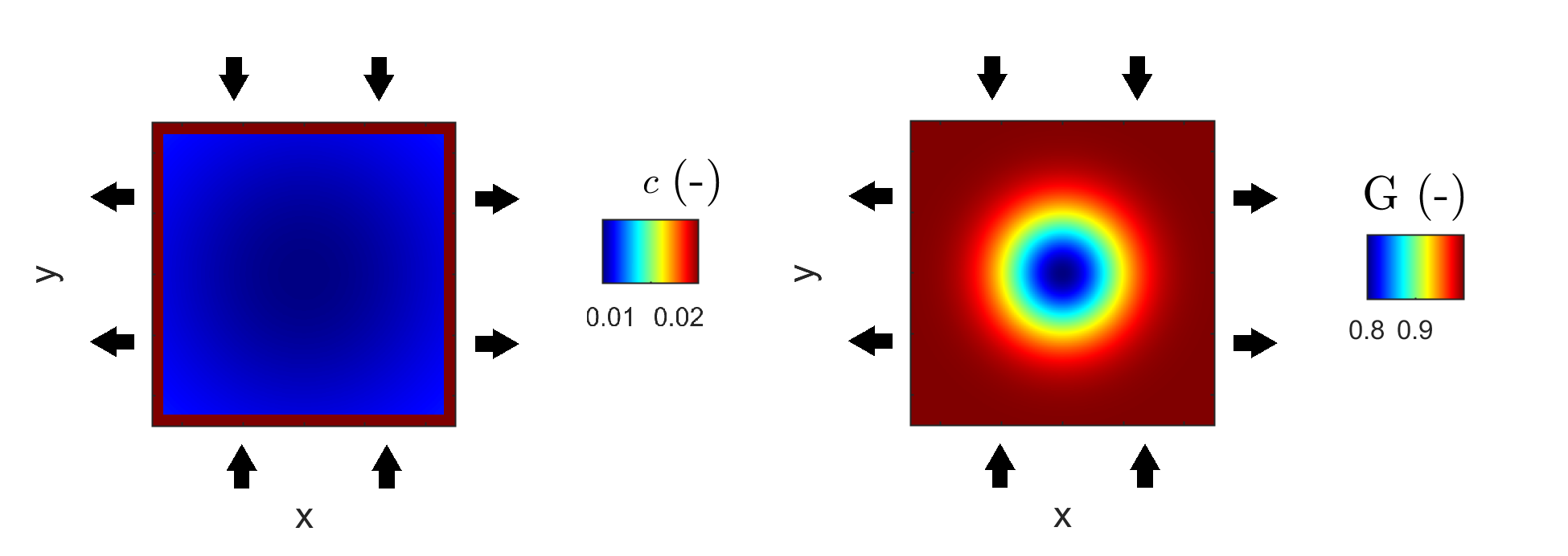}
\caption{Heterogeneous initial setup of cohesion $\bar{c}$ (left) and shear modulus $\bar{G}$ (right). The arrows indicate the pure shear boundary condition which is applied at the model boundaries.} 
\label{Mod12}%
\end{figure}

\begin{figure}
\includegraphics[width=0.7\textwidth]{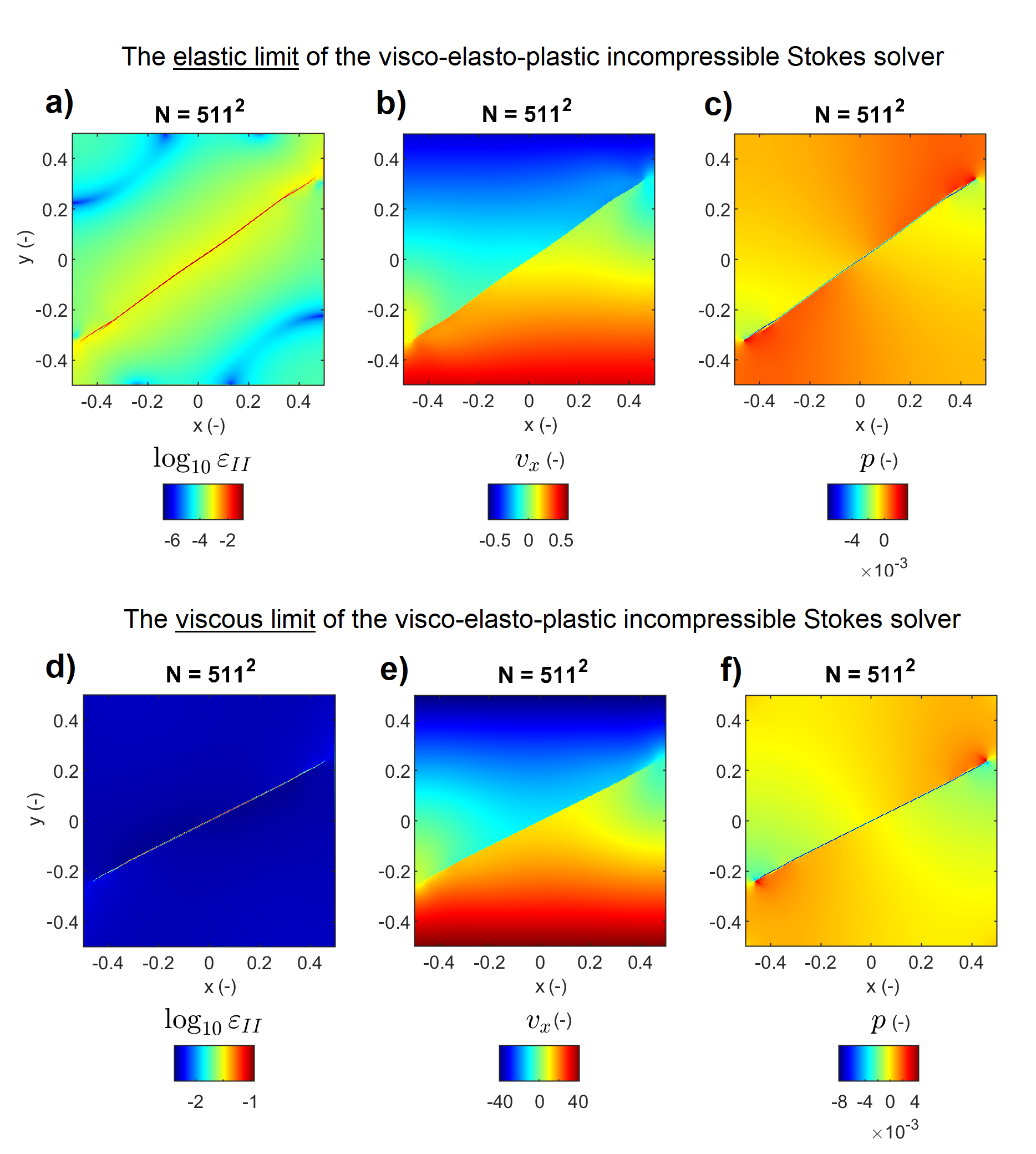}
\caption{Spatial distribution of accumulated strain $\varepsilon_{II}$, velocity $v_x$ and pressure $p$ fields calculated using an incompressible visco-elasto-plastic equations. Panels (a)-(c) correspond to the elastic limit (brittle domain), panels (d)-(f) correspond to the viscous limit (ductile domain). }
\label{ElasticLimit}%
\end{figure}

In the second set of simulations, we explore the incompressible elastic and viscous limits of the incompressible visco-elasto-plastic equations. Figure \ref{Mod12}a shows the initial setup for cohesion, $\bar{c}$; shear modulus, $\bar{G}$, has a constant value of $1$. By setting the Deborah number to the high values $De=10^6$, we reach the elastic limit (brittle domain). Figure \ref{ElasticLimit}a-c shows the accumulated strain $\varepsilon_{II}$, the velocity field in $x$-direction $v_x$ and pressure $p$ for the model configuration corresponding to the elastic limit after 30 loading increments with $\Delta t = 0.4\cdot 10^{-5}$. The model resolution is $N=511^2$ grid cells. The strain localization develops at the center of the model with one initial shear band. The shear band is growing towards the walls of the model as the strain increments evolve in physical time. The velocity field $v_x$ clearly shows a discontinuity along the shear band (Figure \ref{ElasticLimit}b).

%\begin{figure}
%\centering
%\includegraphics[width=0.5\textwidth]{Mod1.png}
%\caption{Heterogeneous initial setup of cohesion $c$. The arrows indicate the pure shear boundary condition which is applied at the model boundaries.} 
%\label{Mod1}%
%\end{figure}

%The localized shear band is distributed over a thickness of a single cell.

Alternatively, by setting the Deborah number to the low value $De=10^{-4}$, the viscous limit is reached (ductile domain). Figure \ref{ElasticLimit}d-f shows the accumulated strain $\varepsilon_{II}$,  the velocity field in $x$-direction $v_x$ and pressure $p$  after 10 loading increments applied to velocity with $a = 1.01$. As in the elastic limit, the velocity field $v_x$ exhibits a discontinuity along the shear band (Figure \ref{ElasticLimit}e). The behavior of the strain localization evolution is similar in both, elastic and viscous, regimes. The only difference between regimes is that in the elastic domain the strain increments are prescribed (equivalent to increments in displacement) whereas in the viscous domain velocity increments are prescribed.

\subsection{Targeting high resolution computations }

\begin{figure}
\centering
\includegraphics[width=0.65\textwidth]{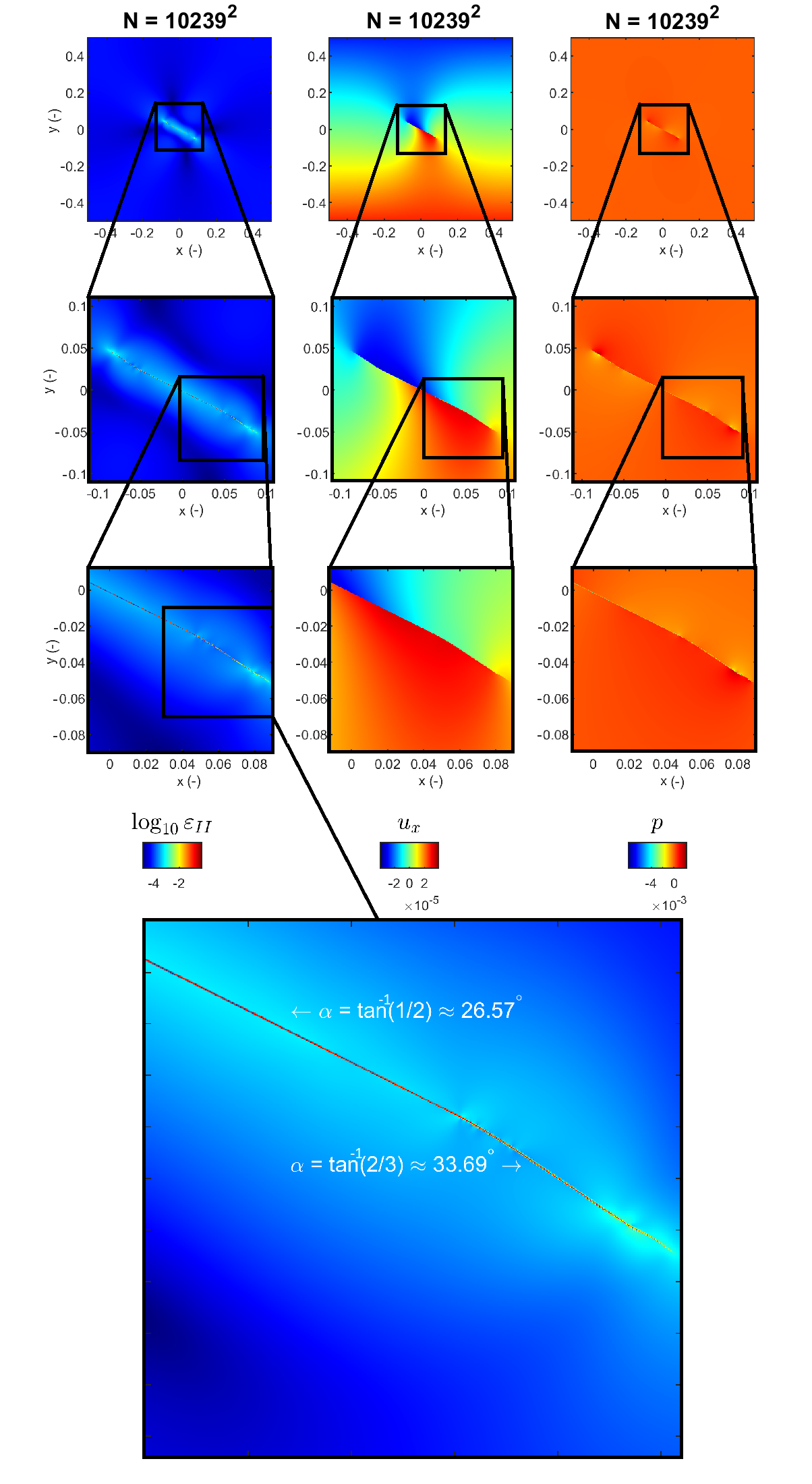}
\caption{Spatial distribution of accumulated strain $\varepsilon_{II}$, displacement $u_x$ and pressure $p$ fields calculated using compressible visco-elasto-plastic equations. The model resolution is $N=10239^2 \approx 105. 000. 000$ grid cells.}
\label{10000}%
\end{figure}

Figure \ref{10000} shows the simulation results of compressible visco-elasto-plastic equations in the brittle domain after 10 loading increments with $\Delta t = 0.2\cdot 10^{-5}$. The model resolution is $N=10239^2 \approx 105. 000. 000$ grid cells. Figure \ref{Mod12} shows the initial setup for cohesion, $\bar{c}$, and shear modulus, $\bar{G}$. The accumulated strain $\varepsilon_{II}$,  the displacement in $x$-direction $u_x$ and pressure $p$ are shown. Similarly to the previous simulations of incompressible equations, a single shear band can be observed. However, the orientation is different; according to our numerical experiments, the orientation might be different for any model configuration. It is clearly visible, that the shear band grows under two different angles with a sharp discontinuity between them. %It means that at initial stage it converges to a different solution and then changes its orientation.

\clearpage

\subsection{Stress drops in two-dimensional simulations}

\subsubsection{Convergence tests for a single stress drop}

%Figure \ref{Temp_conv} shows the integrated stress $\sigma_{xx}$ versus time (Figure \ref{Temp_conv}ae) and the accumulated strain $\varepsilon_{II}$ (Figure \ref{Temp_conv}bcdfgh) for different temporal and spatial resolutions. While the general pattern of the accumulated strain is similar in all simulations (Figure \ref{Temp_conv}bcdfgh), some important features are explored below.

\begin{figure}
\centering
\includegraphics[width=0.99\textwidth]{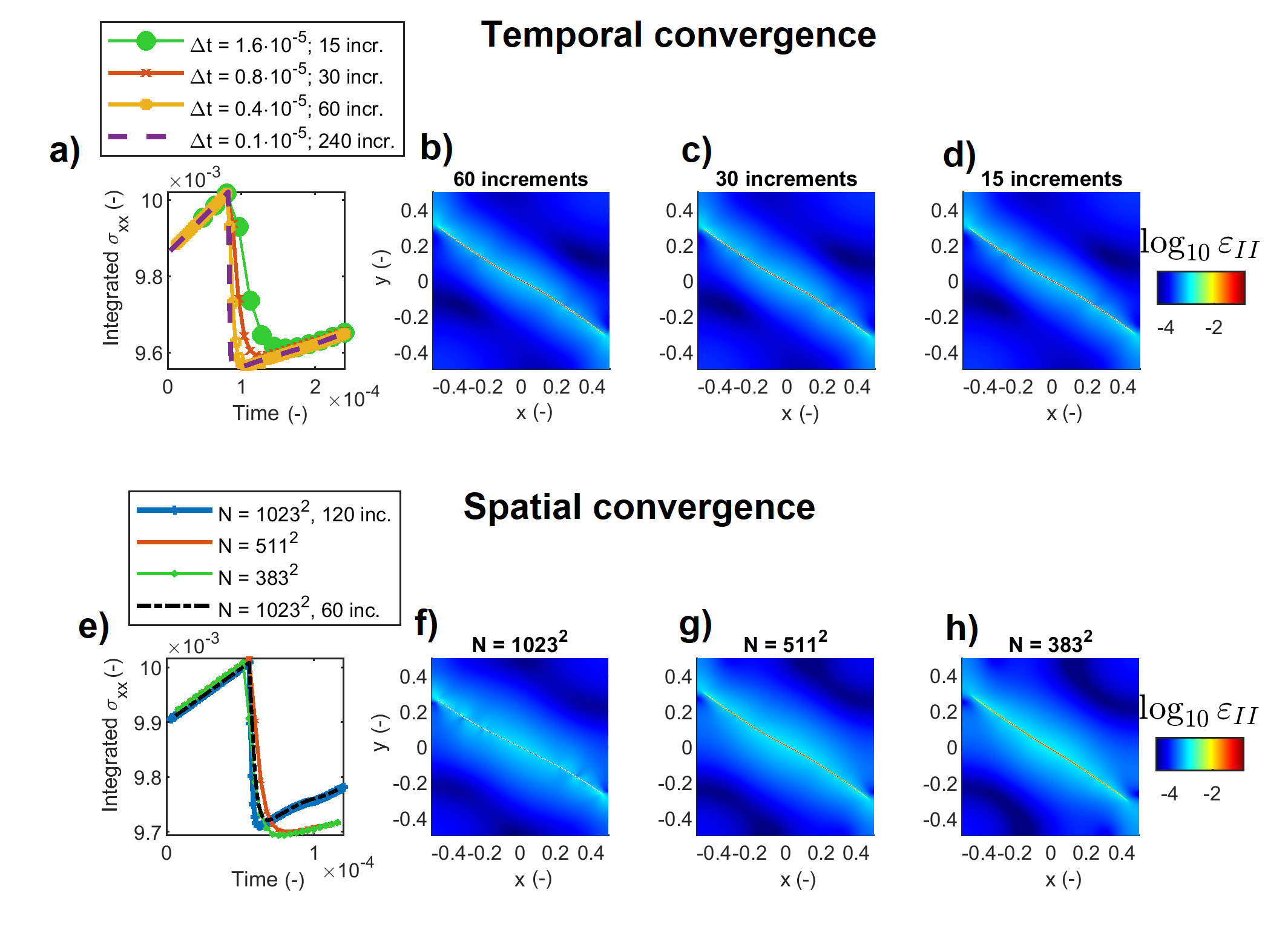}
\caption{ Convergence tests for a single stress drop. Panels (a) and (e) show the integrated stress $\sigma_{xx}$ versus time, panels (b-d) show the accumulated strain $\varepsilon_{II}$ for a set of different temporal discretizations ($N = 511^2 $ grid cells) and panels (f-h) show the accumulated strain $\varepsilon_{II}$ for a set of different spatial discretizations.}
\label{Temp_conv}%
\end{figure}

\begin{figure}
\centering
\includegraphics[width=0.5\textwidth]{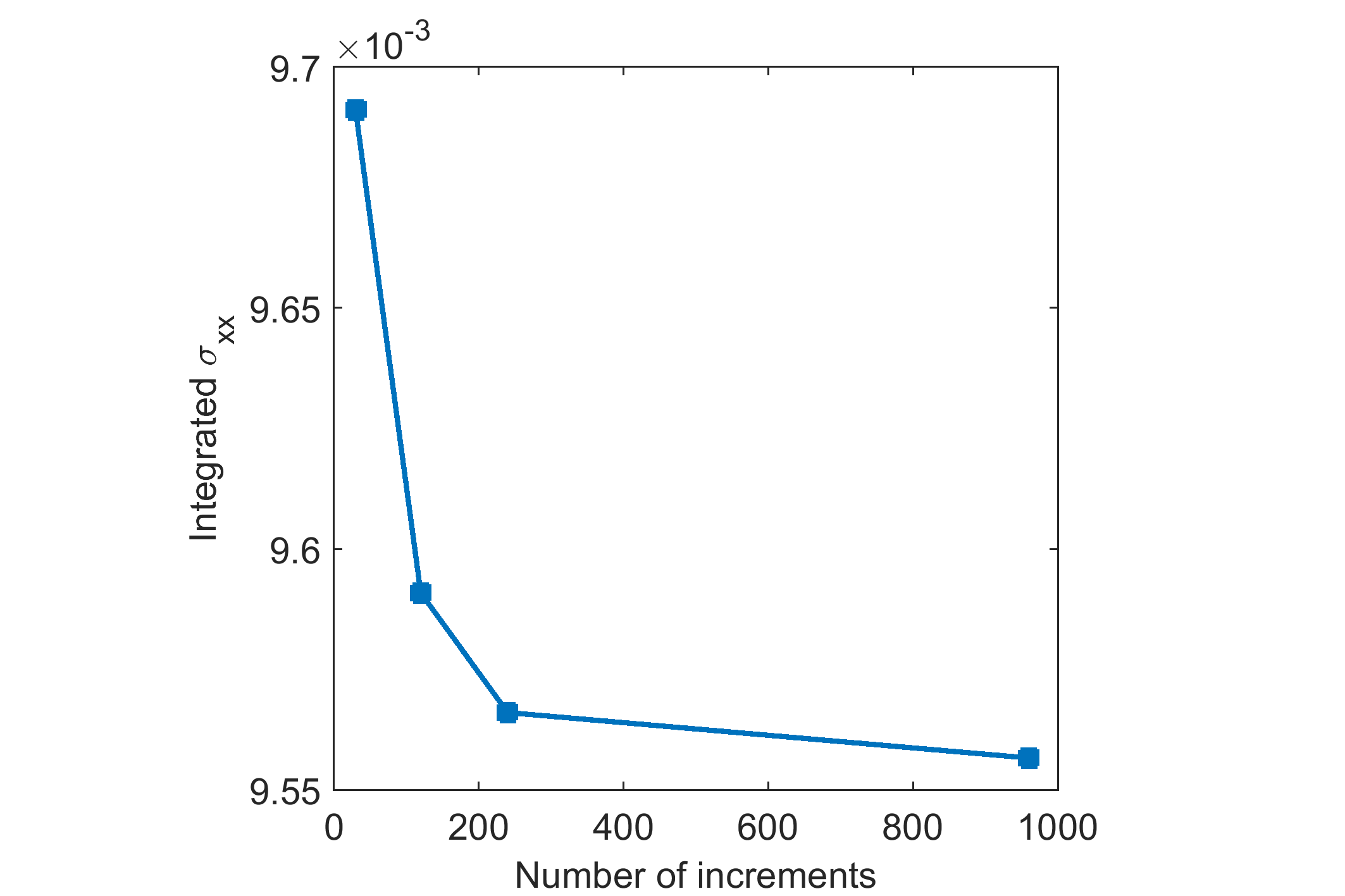}
\caption{The minimum of the integrated stress $\sigma_{xx}$ versus the number of loading increments in time.}
\label{Temp_conv3}%
\end{figure}

First, we run a temporal resolution test (Figures \ref{Temp_conv}a-d). The spatial resolution is $N = 511^2 $ grid cells, simulations with 15, 30, 60, 240 loading increments are performed. Figure \ref{Mod12} shows the initial setup for cohesion, $\bar{c}$, and shear modulus, $\bar{G}$. Note that the integrated stress $\sigma_{xx}$ evolution with loading increments is different depending on the temporal discretization (Figure \ref{Temp_conv}a). The simulations with finer temporal discretization lead to sharper drops in $\sigma_{xx}$ and lower minimum values of stress $\sigma_{xx}$. To further investigate the convergence, we plot the minimum of $\sigma_{xx}$ versus the number of increments in time (Figure \ref{Temp_conv3}). As a result, the minimum of $\sigma_{xx}$ converges to a constant value as temporal resolution increases.

%\subsubsection{Spatial resolution test}
Figure \ref{Temp_conv}e-h shows the integrated stress $\sigma_{xx}$ versus time (Figure \ref{Temp_conv}a) and the accumulated strain $\varepsilon_{II}$ for a set of different spatial discretizations of $N=1023^2$, $N=511^2$ and $N=383^2$ grid cells (Figure \ref{Temp_conv}f-h). A difference in the integrated stress $\sigma_{xx}$ evolution is visible. Note, that the minimum value of the integrated stress $\sigma_{xx}$ is similar for all resolutions.

\begin{figure}
\centering
\includegraphics[width=0.99\textwidth]{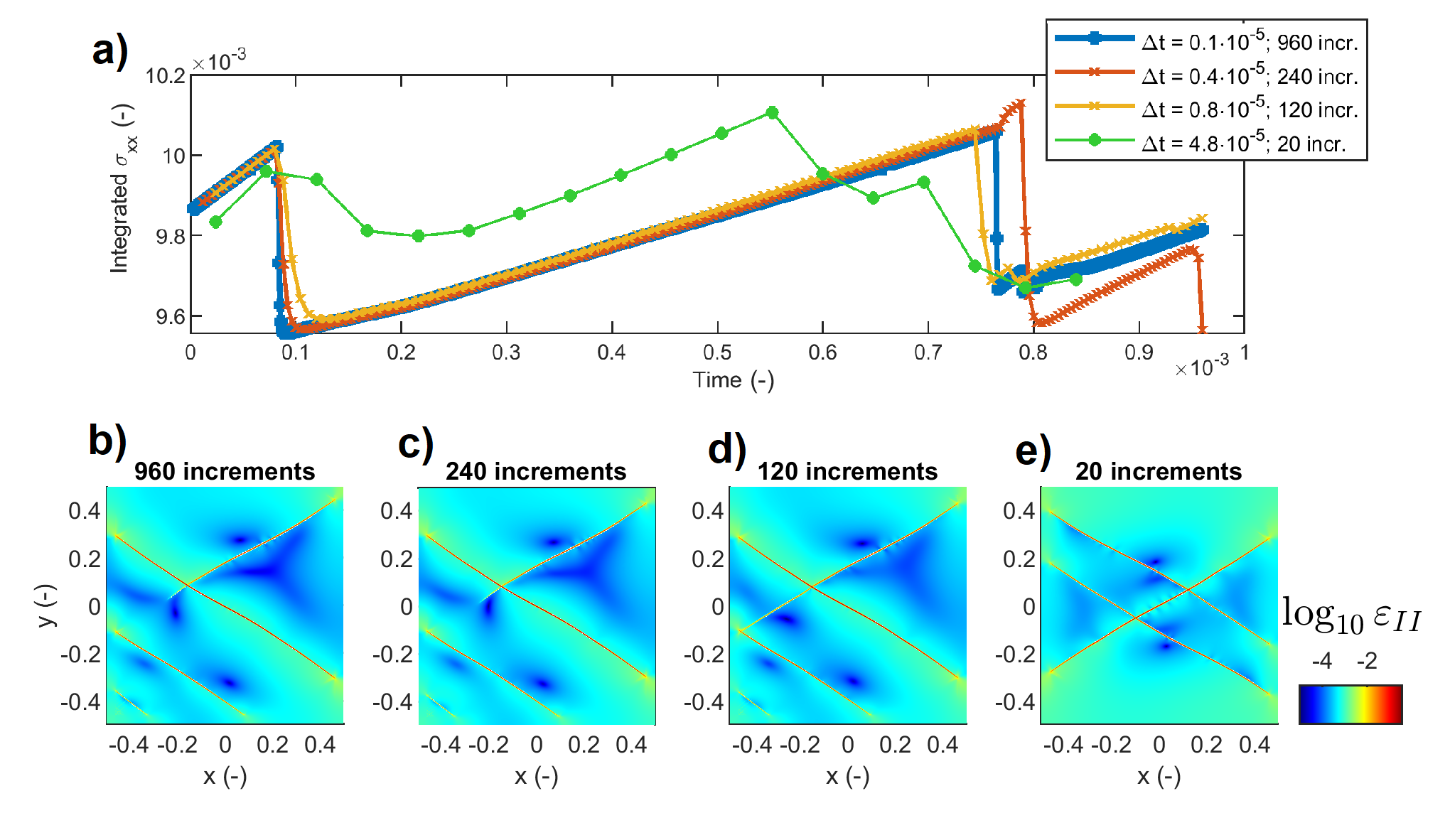}
\caption{Temporal resolution convergence test. Panel (a) shows the integrated stress $\sigma_{xx}$ versus time (loading increments), panels (b-e)  show the accumulated strain $\varepsilon_{II}$ for a set of different temporal discretizations.}
\label{Temp_conv2}%
\end{figure}

To further investigate the dependence of the integrated stress $\sigma_{xx}$ versus temporal resolution, a number of longer simulations were conducted where the second stress drop is visible (Figure \ref{Temp_conv2}). The evolution of the integrated stress $\sigma_{xx}$ around the second stress drop slightly depends of the temporal resolution, however, the difference between simulations of fine resolution is small. The simulation with large loading increments (green curve) significantly diverges from other simulations and don't show stress drops.

%\begin{figure}
%\centering
%\includegraphics[width=0.99\textwidth]{0Spat_conv.png}
%\caption{The spatial resolution numerical test. Panel (a) shows the integrated stress $\sigma_{xx}$ versus increments in time, panels (b-d)  show the accumulated strain ($\varepsilon_{II}$) a set of different spatial discretizations.}
%\label{Sp_conv}%
%\end{figure}

\subsubsection{Stress drop sequence}

\begin{figure}
\centering
\includegraphics[width=0.7\textwidth]{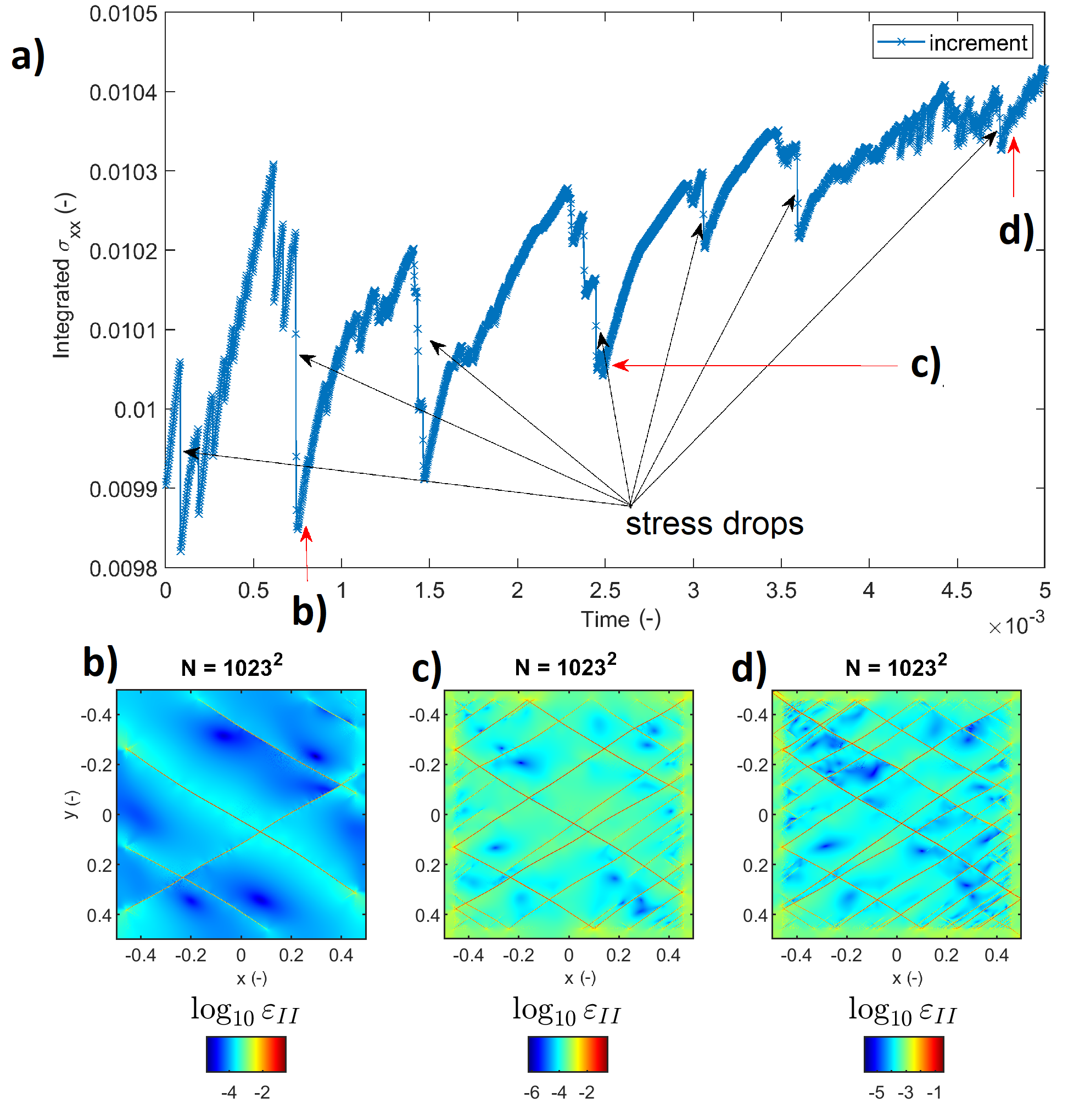}
\caption{Numerical simulation of compressible visco-elasto-plastic equations with the resolution of $1023^2 $ grid cells and $\Delta t = 0.2\cdot 10^{-5}$ for $2500$ loading increments in time. Panel (a) shows the integrated stress $\sigma_{xx}$ versus time. Panels (b-d) show the accumulated strain $\log_{10} \varepsilon_{II}$ for three different loading steps as shown in panel (a) by red arrows.}
\label{Sp_conv2dd}%
\end{figure}

Figure \ref{Sp_conv2dd} shows the integrated stress $\sigma_{xx}$ versus time (Figure \ref{Sp_conv2dd}a) and the accumulated strain $\varepsilon_{II}$ at three discrete loading steps (highlighted by red arrows in Figure \ref{Sp_conv2dd}a). Stress drops have different magnitudes and show non-regular spacing. Stress drops are associated with intersections of shear bands and/or also take place when shear band reaches the higher cohesion values near the boundaries of the model domain.

\subsubsection{Interseismic period and stress drops }

\begin{figure}
\centering
\includegraphics[width=0.7\textwidth]{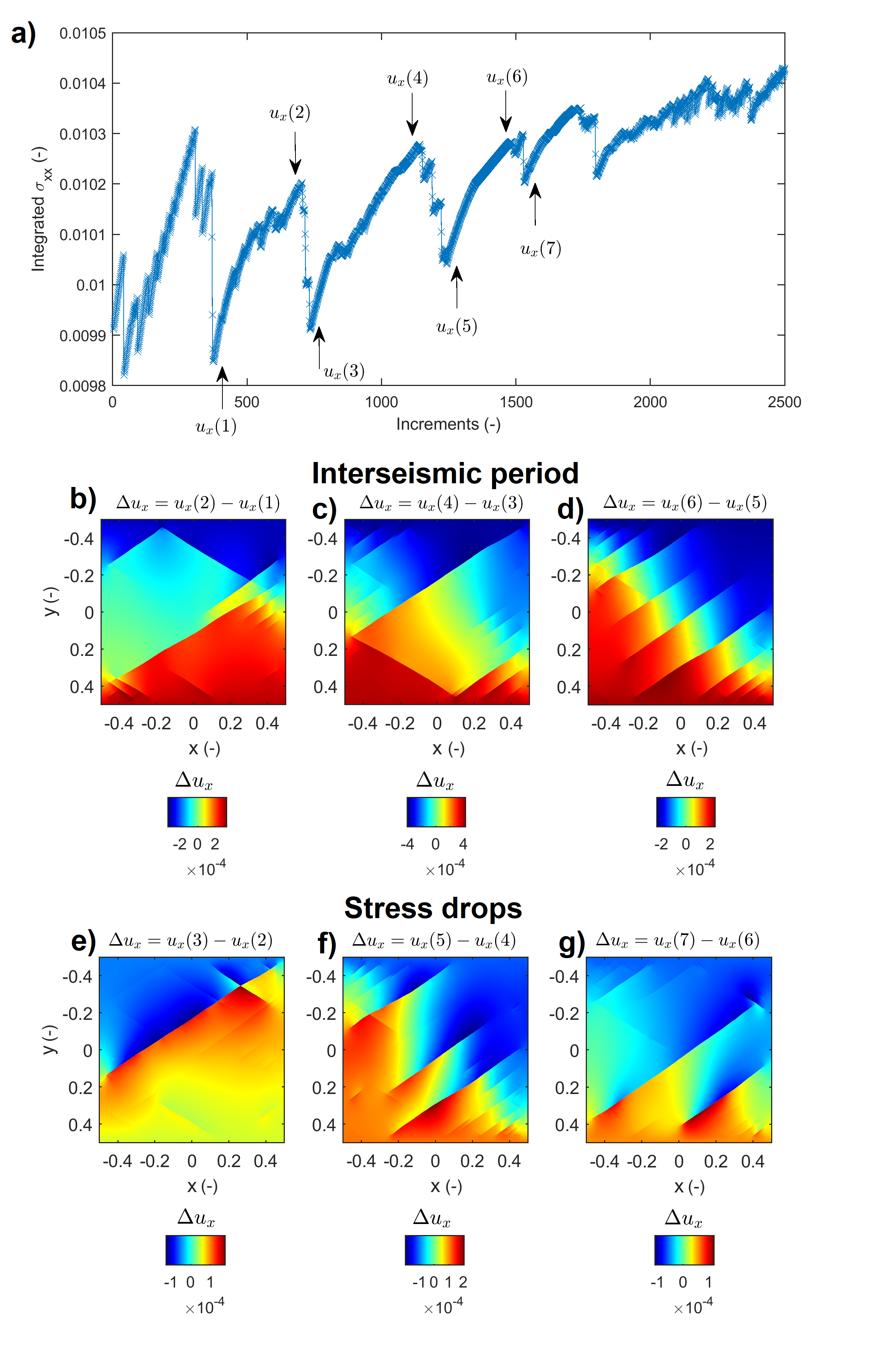}
\caption{Interseismic period and stress drops: numerical simulation of compressible visco-elasto-plastic equations with the resolution of $N = 1023^2 $ grid cells and $\Delta t = 0.2\cdot 10^{-5}$ for $2500$ loading increments in time. Panel (a) shows the integrated stress $\sigma_{xx}$ versus loading increments. Panels (b-d) show displacement increments $\Delta u_x$ corresponding to interseismic periods. Panels (e-g) show displacement increments $\Delta u_x$ corresponding to major stress drops.}
\label{INTERSEISMIC_v1sc}%
\end{figure}

Figure \ref{INTERSEISMIC_v1sc} shows the integrated stress $\sigma_{xx}$ versus loading increments (Figure \ref{INTERSEISMIC_v1sc}a) and the displacement increments $\Delta u_x$ corresponding to the interseismic period (Figure \ref{INTERSEISMIC_v1sc}b-d) and to the stress drops (Figure \ref{INTERSEISMIC_v1sc}e-g). For example, Figure \ref{INTERSEISMIC_v1sc}b shows the displacement increment $\Delta u_x=u_x(2)-u_x(1)$, where $u_x(1)$ and $u_x(2)$ are the displacement fields at the beginning and at the end of the interseismic period (the period between two high-amplitude stress drops, see Figure \ref{INTERSEISMIC_v1sc}a). Figure \ref{INTERSEISMIC_v1sc}cd are similar and correspond to different interseismic periods. In a similar way, the displacement increment $\Delta u_x=u_x(2)-u_x(1)$ of major stress drops are shown (Figure \ref{INTERSEISMIC_v1sc}e-g). It can be seen that displacement accumulates during the interseismic period (without major stress drops) and displacement also accumulates during major stress drops but with increased intensity.

\subsubsection{Earthquake nucleation due to a single stress drop}

\begin{figure}
\centering
\includegraphics[width=0.7\textwidth]{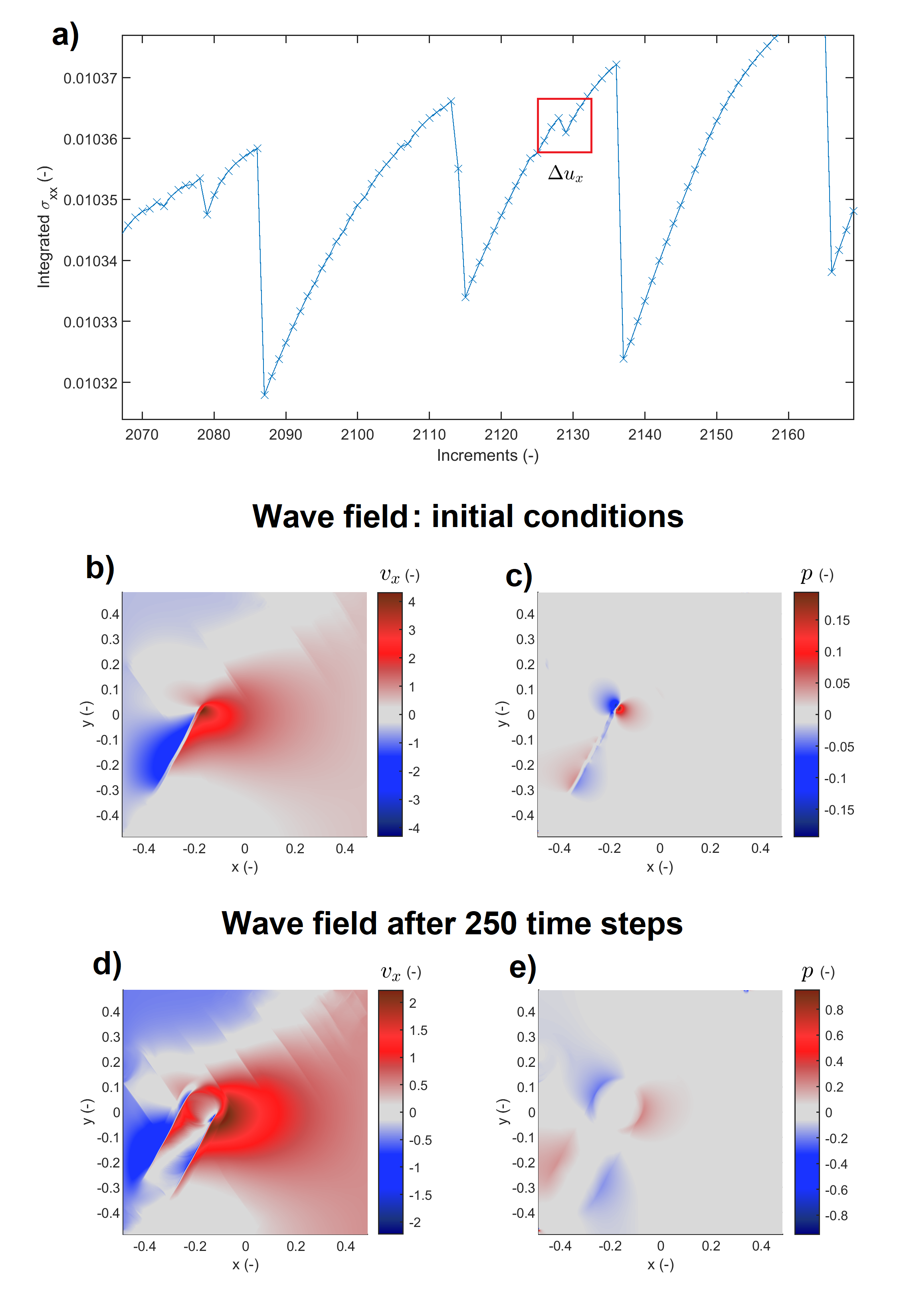}
\caption{Earthquake nucleation due to a single stress drop. Panel (a) shows the integrated stress $\sigma_{xx}$ versus loading increments. Panels (b-c) show the wave fields (velocity $v_x$ and pressure $p$) at the initial stage. Panels (d-e) show wave fields (velocity $v_x$ and pressure $p$) after 250 time steps.}
\label{Wave_P0}%
\end{figure}

Figure \ref{Wave_P0} shows the integrated stress $\sigma_{xx}$ versus loading increments (from 2070 to 2170, Figure \ref{Wave_P0}a)  and the wave fields (velocity $v_x$ and pressure $p$) at the initial stage (Figure \ref{Wave_P0}b-c) and after 250 physical time steps (Figure \ref{Wave_P0}d-e). The wavefield pattern at the initial stage is complex and requires further analysis. The volumetric response is of low amplitude (Figure \ref{Wave_P0}ce) while the velocity field produce high amplitudes (Figure \ref{Wave_P0}bd) meaning that shear component exhibits high amplitudes.

\clearpage

\subsection{Stress drops in three-dimensional simulations}

\subsubsection{Convergence tests for a single stress drop}

\begin{figure}
\centering
\includegraphics[width=0.55\textwidth]{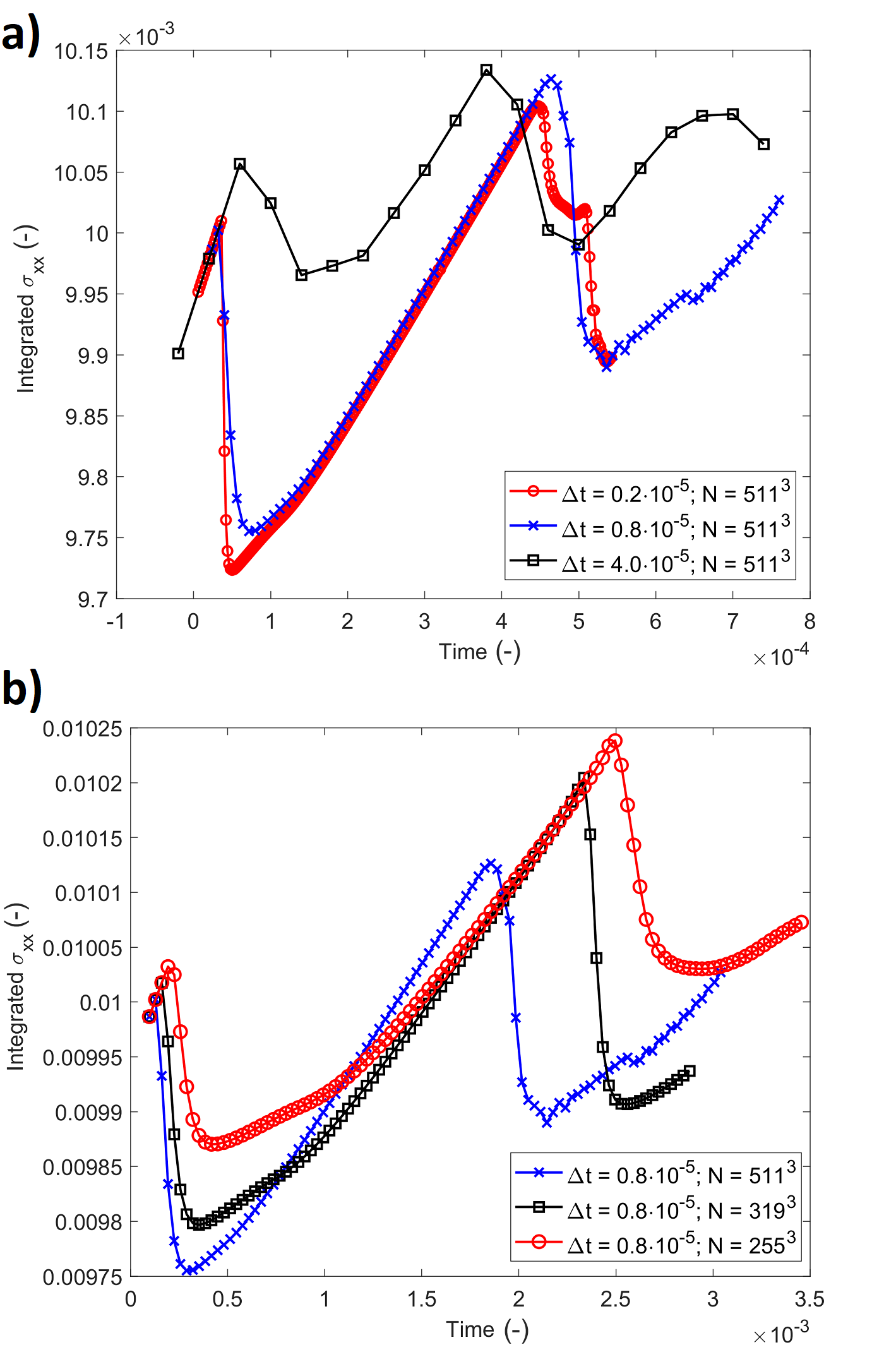}
\caption{Panel (a) --- integrated stress $\sigma_{xx}$ versus time of a model with the resolution of $511^3\approx133. 432. 831$ grid cells for a set of different temporal discretizations ($\Delta t = 0.2\cdot 10^{-5}$, $\Delta t = 0.8\cdot 10^{-5}$ and $\Delta t = 4.0\cdot 10^{-5}$). Panel (b) --- integrated stress $\sigma_{xx}$ versus time of a model with different spatial discretizations ($\Delta t = 0.8\cdot 10^{-5}$).}
\label{03d_conv_v2}%
\end{figure}

Figure \ref{03d_conv_v2}a shows the integrated stress $\sigma_{xx}$ versus time of a model with the resolution of $511^3\approx133. 432. 831$ grid cells for a set of different temporal discretizations ($\Delta t = 0.2\cdot 10^{-5}$, $\Delta t = 0.8\cdot 10^{-5}$ and $\Delta t=4.0\cdot 10^{-5}$). Similarly to the two-dimensional results, the integrated stress $\sigma_{xx}$ evolution with loading increments is different depending on the temporal discretization. The simulations with finer temporal discretization ($\Delta t = 0.2\cdot 10^{-5}$, $\Delta t = 0.8\cdot 10^{-5}$) are similar and lead to sharp drops in $\sigma_{xx}$ and lower minimum values of stress $\sigma_{xx}$. The simulation with coarse temporal discretization ($\Delta t = 4.0\cdot 10^{-5}$) exhibits completely different evolution of the integrated stress $\sigma_{xx}$, which means that the temporal discretization plays an important role in the accuracy of the numerical simulations. Similarly to the two-dimensional results, the minimum of $\sigma_{xx}$ converges to a constant value as temporal resolution increases. Figure \ref{03d_conv_v2}b shows the integrated stress $\sigma_{xx}$ versus time for a set of different spatial discretizations ($\Delta t = 0.8\cdot 10^{-5}$). The simulations with finer spatial discretizations lead to lower minimum values of stress $\sigma_{xx}$. As in two-dimensional examples, a fine spatial and temporal resolutions are needed to accurately describe the stress drop and to evaluate the minimum values of stress $\sigma_{xx}$ corresponding to stress drops.

%Figure \ref{03d_conv_v2} shows the integrated stress $\sigma_{xx}$ versus increments in time (Figure \ref{03d_conv_v2}a) and the accumulated strain ($\varepsilon_{II}$) for a set of different temporal discretizations.

\subsubsection{Stress drops sequence}

\begin{figure}
\centering
\includegraphics[width=0.6\textwidth]{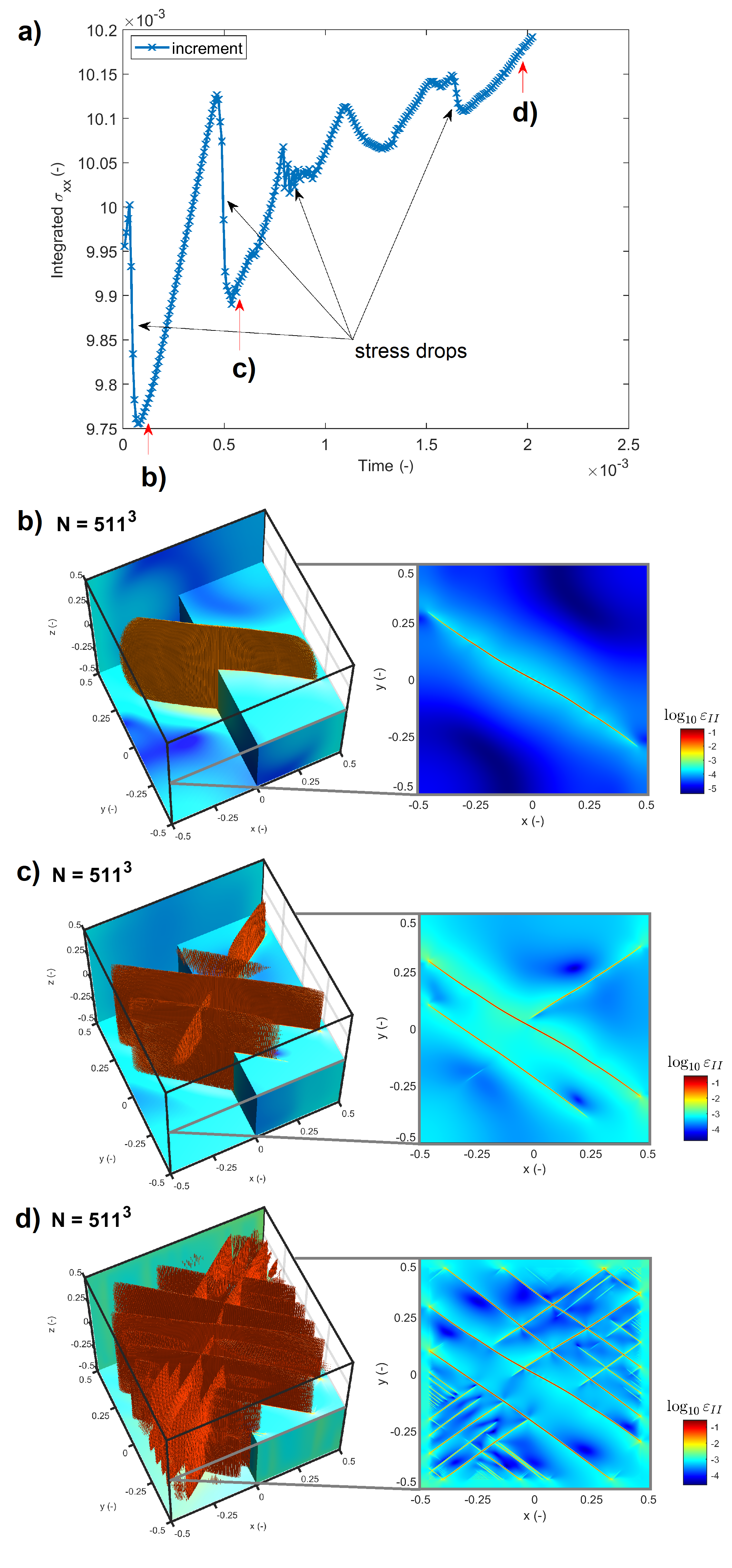}
\caption{Numerical simulation of compressible visco-elasto-plastic equations with the resolution of $511^3 $ grid cells and $\Delta t = 0.8\cdot 10^{-5}$ for $250$ increments in time. Panel (a) shows the integrated stress $\sigma_{xx}$ versus increments in time; panels (b-d) show the accumulated strain $\log_{10} \varepsilon_{II}$ for three different times as shown in panel (a).}
\label{Sp_conv3dd}%
\end{figure}

Figure \ref{Sp_conv3dd} shows the results of a numerical simulation with the resolution of $511^3\approx133. 432. 831$ grid cells grid cells and $\Delta t = 0.8\cdot 10^{-5}$ for $250$ loading increments in time. In total, four sharp stress drops in the integrated stress $\sigma_{xx}$ are visible; the stress drops magnitudes are also different. The place and magnitude of the stress drops is spontaneous as in the two-dimensional simulations.

\section{Discussion}
 
There are three main numerical parameters which are crucial for the proper numerical simulations of strain localization: a high spatial resolution, a high temporal resolution and the convergence of iterations at each loading increment to the precision of $err=10^{-12}$. If one of these three conditions is not satisfied --- the numerical results might be inaccurate and show a symmetric pattern. According to our tests, a resolution of $500\times500$ grid cells or more is usually sufficient to resolve the strain localization pattern. To analyze the evolution of the angle of shear bands, a resolution of $1000\times1000$ or more is needed. However, only small loading increments made it possible to observe sharp stress drops. Sharper stress drops are observed in simulations with smaller loading increments (Figures \ref{Sym_191v3S}g, \ref{Temp_conv2}a, \ref{03d_conv_v2}a).

\clearpage

\section{Conclusions}

Numerical simulations have been performed addressing different aspects of the modeling of incompressible and compressible visco-elasto-plastic inertialess equations. We have shown that the strain localization pattern as well as velocity discontinuity are equally presented in both elastic and viscous limits of the incompressible visco-elasto-plastic equations. We have presented a high resolution two-dimensional simulation of $\approx 10000^2$ grid cells and showed that the shear band grows under two different angles during its evolution. We concluded that there are three main numerical parameters which are crucial for the proper numerical simulations of strain localization: a high spatial resolution, a high temporal resolution and the convergence of iterations at each loading increment to the precision of $\epsilon_\mathrm{rel}=10^{-12}$. If one of these three conditions is not satisfied --- the numerical results might be inaccurate. We have investigated the integrated stress of the model at each loading increment and found that it exhibit sudden stress drops. Such stress drops are spontaneous and exhibit different magnitudes. We have analyzed the spatial and temporal convergence of two- and three- dimensional simulations with the spatial resolution of $\approx 1000^2$ and $\approx 500^3$ grid cells, respectively. Large two- and three-dimensional simulations leading to multiple stress drops are also presented. Such stress drops may correspond to the sequence of earthquakes, therefore, further research is needed.

%\pagebreak

%\pagebreak

\newpage

\newpage

\appendix
\addcontentsline{toc}{section}{Appendices}
\section{The objective stress terms}\label{objective_stress}

After expanding, collecting and rearranging terms, the objective stress terms $\tau_{ij}^{\mathcal{R}}$ for a three-dimensional configuration are (in Voigt notation)
\begin{align}	
		\sigma_{xx}^{\mathcal{R}}&= 2(\sigma_{xy}\dot{\omega}_{xy}+\sigma_{xz}\dot{\omega}_{xz})\label{sigxxR},\\
		\sigma_{yy}^{\mathcal{R}}&=-2(\sigma_{xy}\dot{\omega}_{xy}-\sigma_{yz}\dot{\omega}_{yz})\label{sigyyR},\\
		\sigma_{zz}^{\mathcal{R}}&=-2(\sigma_{xz}\dot{\omega}_{xz}+\sigma_{yz}\dot{\omega}_{yz})\label{sigzzR},\\
		\sigma_{xy}^{\mathcal{R}}&=\dot{\omega}_{xy}(\sigma_{yy}-\sigma_{xx})+\sigma_{yz}\dot{\omega}_{xz}+\sigma_{xz}\dot{\omega}_{yz}\label{sigxyR},\\
		\sigma_{yz}^{\mathcal{R}}&=\dot{\omega}_{yz}(\sigma_{zz}-\sigma_{yy})-\sigma_{xy}\dot{\omega}_{xz}-\sigma_{xz}\dot{\omega}_{xy}\label{sigyzR},\\
		\sigma_{xz}^{\mathcal{R}}&=\dot{\omega}_{xz}(\sigma_{zz}-\sigma_{xx})+\sigma_{yz}\dot{\omega}_{xy}-\sigma_{xy}\dot{\omega}_{yz}\label{sigxzR},
	\end{align}
and, for a two-dimensional configuration assuming plane strain conditions, Eqs. \ref{sigxxR}, \ref{sigyyR} and \ref{sigxyR} reduce to
\begin{align}
		\sigma_{xx}^{\mathcal{R}}&=2\sigma_{xy}\dot{\omega}_{xy},\\
		\sigma_{yy}^{\mathcal{R}}&=-2\sigma_{xy}\dot{\omega}_{xy},\\
		\sigma_{xy}^{\mathcal{R}}&=\dot{\omega}_{xy}(\sigma_{yy}-\sigma_{xx}).
	\end{align}%($\sigma_{zz}=\sigma_{yz}=\sigma_{zx}=0$)

\section*{Acknowledgements}
%%%%%%%%
Yury Alkhimenkov gratefully acknowledges support from the Swiss National Science Foundation, project number P500PN$\_$206722. Authors gratefully acknowledge support from the Ministry of Education and Science of Russian Federation (grant №075-15-2022-1106).

\section*{Author contributions}
\textbf{YA}: Conceptualization, Methodology, Software, Writing -- Original Draft, Visualization, Investigation, Formal analysis, Project administration. \textbf{LK}: Methodology, Software, Writing -- review \& editing. \textbf{IU}: Methodology, Software, Writing -- review \& editing. \textbf{YP}: Conceptualization, Methodology, Software, Writing -- review \& editing, Supervision.

\section*{Data Availability Statement}
No data were used in producing this manuscript.

  \clearpage

%\end{thebibliography}
\bibliographystyle{apa}
\bibliography{agusample}

\end{document}